\documentclass[11pt,onecolumn]{article}
\usepackage{ISG_style}
\usepackage{times}
\usepackage{graphicx,subfigure}
\usepackage{color}
\usepackage{amsmath,amssymb} 
\usepackage{dsfont}
\usepackage{cite}
\usepackage{hyperref}
\usepackage{epstopdf}
\usepackage{url}

\usepackage{algorithm}
\usepackage{makecell}
\usepackage{algcompatible}
\usepackage{algpseudocode}
\usepackage{mathtools}
\usepackage{verbatim}
\usepackage[font=small,justification=centering]{caption}
\ifpdf
    \graphicspath{{figures/PNG/}{figures/PDF/}{figures/}}
\else
    \graphicspath{{figures/EPS/}{figures/}}
\fi

\DeclareMathOperator*{\argmax}{arg\,max}  

\newcommand{\filter}{{\sf H}}

\newcommand{\grnn}{{\sf GRNN}}
\newcommand{\grnns}{{\sf GRNNs}}
\newcommand{\gnn}{{\sf GNN}}

\newcommand{\gcnn}{{\sf GCNN}}
\newcommand{\gcnns}{{\sf GCNNs}}

\newcommand{\nn}{{\sf NN}}
\newcommand{\nns}{{\sf NNs}}
\newcommand{\fcs}{{\sf FCs}}
\newcommand{\fc}{{\sf FC}}
\newcommand{\dc}{{\sf DC}}
\newcommand{\ac}{{\sf AC}}

\newcommand{\mws}{{\sf MWs}}

\newcommand{\GRQN}{{\sf GRQN}}
\newcommand{\GRQNs}{{\sf GRQNs}}

\newcommand{\ml}{{\sf ML}}
\newcommand{\rl}{{\sf RL}}
\newcommand{\PFW}{{\sf PFW}}
\newcommand{\TE}{{\sf TE}}
\newcommand{\mdp}{{\sf MDP}}

\newcommand{\pomdp}{{\sf POMDP}}
\newcommand{\pomdps}{{\sf POMDPs}}
\newcommand{\episode}{{\sf Episode}}
\newcommand{\buffer}{{\sf Buffer}}
\newcommand{\load}{{\sf load}}
\newcommand{\loss}{{\sf LL}}
\newcommand{\tll}{{\sf TLL}}
\newcommand{\isend}{{\sf end}}
\newcommand{\pf}{{\sf PF}}
\newcommand{\mc}{{\sf MC}}

\title{\bf \LARGE GRNN-based Real-time Fault Chain Prediction}

\date{}

\author{Anmol~Dwivedi and Ali~Tajer \\ 
Electrical, Computer, and Systems Engineering Department\\
Rensselaer Polytechnic Institute} 
\begin{document}

\maketitle

\begin{abstract}
    This paper proposes a data-driven graphical framework for the real-time search of risky cascading fault chains ($\fcs$). While identifying risky $\fcs$ is pivotal to alleviating cascading failures, the complex spatio-temporal dependencies among the components of the power system render challenges to modeling and analyzing $\fcs$. Furthermore, the real-time search of risky $\fcs$ faces an inherent combinatorial complexity that grows exponentially with the size of the system. The proposed framework leverages the recent advances in graph recurrent neural networks to circumvent the computational complexities of the real-time search of $\fcs$. The search process is formalized as a partially observable Markov decision process ($\pomdp$), which is subsequently solved via a time-varying graph recurrent neural network ($\grnn$) that judiciously accounts for the inherent temporal and spatial structures of the data generated by the system. The key features of this structure include (i) leveraging the spatial structure of the data induced by the system topology, (ii) leveraging the temporal structure of data induced by system dynamics, and (iii) efficiently summarizing the system's history in the latent space of the $\grnn$. The proposed framework's efficiency is compared to the relevant literature on the IEEE $39$-bus New England system~{and the IEEE $118$-bus system}.
\end{abstract}

\section{Introduction}
\label{sec:Introduction}

Large-scale disruptions in power systems generally follow a sequence of less severe anomalous events that gradually stress the system over time. Various reports on the events preceding blackouts indicate that system operators are either unaware of these gradual changes or oblivious to the contingencies following these changes. Specifically, such small-scale anomalies can lead to hidden failures that propagate and eventually result in large-scale monitoring, and control disruptions. For instance, in a report by the North American Electric Reliability Corporation~\cite{NAPS:2004}, it was concluded that inadvertent tripping of a power line led to a series of failures causing the 2003~blackout in North America. Therefore, forming \emph{real-time and accurate} situational awareness in power systems has a pivotal role in ensuring a secure and reliable power system operation.

In this paper, we formalize a graphical framework for dynamically predicting the chains of risky faults a power system faces. A fault chain $(\fc)$ is a sequence of consecutive component outages that captures the temporal evolution of a cascading outage process. Due to the combinatorially extensive number of possible failures, which grows with the system size and failure horizon, finding critical failure sequences is computationally challenging. Since a $\fc$ captures the high-impact and rear-occurrence characteristic of cascading failures in any dynamically-changing power system partly, timely identification of the riskiest $\fcs$ to system operators for any given system state is instrumental to predicting failures and preventing cascading failures.

Predicting $\fcs$ and assessing their risks can be formalized by (i) creating an exhaustive list of all possible failure scenarios up to a specific time horizon, (ii) evaluating the disruption (e.g., load loss) caused by each, and (iii) evaluating the likelihood of each scenario. Accomplishing these three tasks faces the following two key challenges. First, the space of scenarios grows exponentially fast with the power system size and time horizon, rendering listing all scenarios computationally prohibitive even for moderate network sizes and target horizons. Secondly, the system changes dynamically with high unpredictability, which necessitates constantly updating the scenario space. Before specifying our approach, we review the existing literature relevant to this paper's scope.


\subsection{Literature Review}

In this subsection, we provide an overview of the literature most closely related to the scope of this paper. The study in~\cite{Wang:2011} aims to identify and quantify all the vulnerable sections of the power system that can potentially lead to cascading failures. Specifically, it develops a  $\fc$ framework for risky $\fc$ identification to address this. In~\cite{RandomChemistryModel}, a rapid stochastic procedure is proposed to yield large collections of high-risk $\fcs$. Despite the effectiveness of~\cite{RandomChemistryModel}, such a method faces a computational bottleneck addressed by the study in~\cite{Yang:2017}, which proposes eliminating a large number of redundant constraints to make the contingency screening more efficient. With similar motivation,~\cite{Ding:2017} formulates a bi-level optimization problem to gain a higher evaluation efficiency for risky $\fc$ search and~\cite{Guo:2018} employs a sequential importance sampling algorithm to acquire critical $\fcs$ from cascading outage simulations. Due to a variety of other benefits associated with identifying a collection of risky $\fcs$, $\fc$ search algorithms are employed for risk-assessment~\cite{Rezaei:2015, Henneaux:2016, Rui:2017}, risk mitigation~\cite{Yao:2018}, and vulnerable component identification~\cite{Wei:2018, Zhang:2020} and others~\cite{Pattern:2021}. 

There exist a number of machine learning $(\ml)$ approaches that aim to enhance the efficiency of risky $\fc$ search. Broadly, these models quantify the vulnerability of each power system component from simulated power system operational data by learning data-driven models. For instance, the study in~\cite{Yan:2017} formulates the search for risky $\fcs$ in a Markov decision process ($\mdp$) environment and employs reinforcement learning ($\rl$) algorithms to find risky $\fcs$. The investigation in~\cite{Li:2018} employs deep neural networks to obtain critical states during cascading outages, and~\cite{Du:2019} employs convolutional neural networks for faster contingency screening. The study in~\cite{Zhang:2020} proposes a transition-extension approach that builds upon the $\rl$ approach in~\cite{Yan:2017} to make it amenable to real-time implementation. This is facilitated by exploiting the similarity between adjacent power-flow snapshots. Finally,~\cite{Liu:2021} proposes to employ graph convolutional neural networks that leverage the grid's topology to identify cascading failure paths.

In parallel to ML-based approaches, there also exist model-based approaches that quantitatively and qualitatively model, analyze and simulate large-scale cascading outage processes. This is done by developing simulation models that capture the power system physics. The first category of studies aims to develop high-level statistical models to facilitate faster computation and quick inference. Generally, these are data-driven models such as CASCADE~\cite{Dobson:2003}, the branching process model~\cite{Dobson:2004}, the interaction model~\cite{Sun:2015}, and models based on influence graphs~\cite{Dobson:2017, Wu:2021}. While these approaches are quick in revealing important quantitative properties of cascading outages and often lead to interpretable conclusions due to their ability to capture the non-local behavior of the cascading outage process, such models cannot accommodate the time-varying interactions among components at different stages of a cascade. To tackle this challenge, detailed failure models such as the OPA~\cite{Dobson:OPA}, improved OPA~\cite{Improved:OPA}, random chemistry model~\cite{RandomChemistryModel} and others~\cite{Soltan:2017, Cetinay:2018} are studied that precisely model the $\ac$ power-flow and power dispatch constraints of the components in the grid while simulating the cascading outage process. Despite their detailed modeling and effectiveness, these models are computationally intensive, a major impediment to their adoption for real-time implementation. 

\subsection{Contribution}

Despite the effectiveness of the aforementioned models, these models broadly face the following challenges: (i) The models fail to capture the concurrent spatio-temporal dependencies across the time horizon among the components of the power system under dynamically changing network topologies. (ii) The cascading outage process is assumed to follow Markov property. While this simplification can render reasonable approximations during the earlier stages of the cascading failure, the later stages of the failure process typically exhibit temporal dependencies beyond the previous stage, making the Markovian assumptions inadequate. (iii) These models become prohibitive even for moderate grid sizes due to the combinatorial growth in either the computational or storage requirements with the number of components in the grid. 

We propose a data-driven graphical framework for efficiently identifying risky cascading $\fcs$ to address the aforementioned challenges associated with the $\ml$ approaches. The proposed framework designs a graph recurrent neural network ($\grnn$) to circumvent the computational complexities of the real-time search of $\fcs$. The search process is formalized as a partially observable Markov decision process ($\pomdp$), which is subsequently solved via a time-varying $\grnn$ that judiciously accounts for the inherent temporal and spatial structures of the data generated by the system. The key features of this structure include (i) leveraging the spatial structure of the data induced by the system topology, (ii) leveraging the temporal structure of data induced by system dynamics, and (iii) efficiently summarizing the system's history in the latent space of the $\grnn$, rendering the modeling assumptions realistic and the approach amenable to real-time implementation.  

Finally, we highlight the difference between our proposed approach and the graph neural network ($\gnn$)-based approach investigated in~\cite{Liu:2021}. The goal in~\cite{Liu:2021} is to detect all cascading failure paths that lead to load-shedding in a limited number of search attempts. Such a decision is~\emph{binary} since a cascading failure path may either lead to load shedding or not. In contrast, our approach aims to identify cascading failure paths with~\emph{maximum risk} ($\fcs$ with maximum load-shed) in a limited number of search attempts where, ideally, the most critical cascading failure paths should be identified~\textbf{earlier} than the relatively less critical ones. While related, the problem descriptions and the attendant solutions are distinct.

\section{Problem Formulation}
\label{sec:Problem Formulation}

\subsection{System Model}
\label{sec:System Model}

Consider a power system consisting of $N$ buses. To capture the interconnectivity of the system, we represent it by an undirected graph $\mcG \dff (V, E)$, where the set of vertices $V\dff[N]\dff\{1,\dots,N\}$ represents the buses and the edge set $E\subseteq V\times V$ represents the transmission lines. We denote the binary adjacency matrix of $\mcG$ by $\bB \in \{0,1\}^{N \times N}$ such that $b_{u,v}\dff [\bB]_{u,v}=1$ indicates~{there exists at least one transmission line between buses $u$ and $v$ (e.g., in the case of parallel transmission lines)}. We define $\bX\in\R^{N\times F}$ as the~{system} state matrix of the grid, which compactly represents $F$~{system} state parameters (e.g., voltage angles, injected real power) for all the $N$ buses and we refer to any~{system} state parameter $f\in\{1,\dots,F\}$ of bus $u\in [N]$ by $x_{u,f}\dff [\bX]_{u,f}$. 

\subsection{Modeling Fault Chains}
\label{sec:Modeling Fault Chains}

Depending on an array of internal (e.g., system instabilities) and external (e.g., weather) conditions, the system faces a degree of risk in disruptions that may lead to component outages. We focus on transformers and transmission lines as the components of interest. When the outages are substantial enough, they can lead to more outages, resulting in a $\fc$. Our objective is to dynamically identify the $\fcs$ that the system faces and assess their associated risks (e.g., load losses). In this subsection, we formalize a model for $\fc$s and their risks based on an objective formalized in the next subsection.

Consider the topology of a generic outage-free system that precedes an $\fc$ given by $\mcG_{0}$, and denote the associated system state by $\bX_{0}$. A generic $\fc$ that the system might be facing is specified as a sequence of consecutive component outages that represents a cascading outage process. Consider a $\fc$ model that consists of {at most} $P$ stages, where $P$ can be selected based on the horizon of interest for risk assessment, and denote the set of all components in the system by $\mcU$. {It is noteworthy that the number of stages in different $\fcs$ can be distinct. In cases that a fault chain terminates before $\tilde P< P$ stages, it means that the sets $\{{\cal U}_{\tilde P+1},\dots, {\cal U}_{P}\}$ will be empty sets.}

In each stage $i\in[P]$, a number of additional components fail. We define $\mcU_i$ as the set of components that fail in stage $i\in[P]$. {We note that the set $\mcU_i$ could consist of more than one component failures in any stage $i$, i.e., $|\mcU_i| \geq 1$. We also note that when a transmission line fails and it has parallel channels, only the failed line will be included in $\mcU_i$, and its parallel lines will be retained as functioning components}. Clearly, $\mcU_i\subseteq \mcU\backslash \{\cup_{j=1}^{i-1}\mcU_j\}$, and the sequence of components that fail in $P$ stages is specified by a $\fc$ sequence $\mcV$
\begin{align}
    \mcV\dff \langle \mcU_1, \mcU_2, \dots, \mcU_P \rangle\ .
\end{align} 
Additionally, we denote the healthy components in any stage $i$ of the $\fc$ by $\ell_{i} \in \mcU\backslash \{\cup_{j=1}^{i-1}\mcU_j\}$. We note that when the failure process is slow (e.g., in the earlier stages of a cascading failure), the sets $\mcU_i$ have fewer components, and when the process is fast (e.g., last stages of a cascading failure), these sets are more populated since component outages are usually grouped depending on their time of occurrence~\cite{Henneaux:2016, Rui:2017, Dobson:2017}.

Due to component outages in each stage $i$, the system's topology alters to $\mcG_{i} \dff (V_{i}, E_{i})$ with an associated adjacency matrix $\bB_{i}$ and an underlying system state $\bX_{i}$. Consecutive failures lead to compounding stress on the remaining components, which need to ensure minimal load shedding in the network. Nevertheless, when the failures are severe enough, they can lead to load losses. We denote the load loss $(\loss)$ imposed by the component failures in $\mcU_i$ in stage $i$ by $\loss(\mcU_i)$, i.e., 
 \begin{align}
    \label{eq:loss}
    \loss(\mcU_{i}) \dff \load(\mcG_{i-1}) - \load(\mcG_i)\ ,
\end{align} where $\load(\mcG_{i})$ is the total load (in $\mws$) when the system's state in stage $i$ is associated with the topology $\mcG_i$. Accordingly, we define the total load loss ($\tll$) imposed by the $\fc$ $\mcV$ by
\begin{align}
    \label{eq:risk until given stage}
    \tll(\mcV) \dff \sum_{i=1}^{P}\; \loss(\mcU_{i})\ .
\end{align} 

\subsection{Problem Statement}
\label{sec:Problem Statement}

Due to the re-distribution of power across transmission lines after each stage of the $\fc$, some $\fc$ sequences particularly lead to substantial risks and owing to the continuously time-varying system's state, different loading and topological conditions face different risks. Hence, it is important for system operators to find, efficiently and in real-time, the set of $\fcs$ with the largest $\tll$ associated with any given initial system state $\bX_0$.

Our objective is to identify $S$ {number of} $\fc$ sequences, each consisting of $P$ stages, that impose the largest $\tll$. To formalize this, we define ${\mcF}$ as the set of all possible $\fcs$ with a target horizon of $P$, and our objective is to identify $S$ members of~{$\mcF$} with the largest associated losses. We denote these $S$ members by $\{\mcV^*_1,\dots,\mcV^*_S\}$. Identifying the sets of interest can be formally cast as solving
\begin{align}
    \label{eq:OBJ1}
    \mcP:\quad  \{\mcV^*_1,\dots,\mcV^*_S\} \dff  \argmax_{\{\mcV_1,\dots,\mcV_S\}\;: \mcV_i\in{\mcF}} \;  \sum_{i = 1}^{S} \; \tll(\mcV_{i})\ . 
\end{align} 
Problem $\mcP$ aims to maximize the accumulated $\tll$ due to the $S$ {number of} $\fc$ sequences. Without loss of generality, we assume that the $\tll$s of the set of sequences $\{\mcV^*_1,\dots,\mcV^*_S\}$ are in the descending order, i.e., $\tll(\mcV^*_1) \geq \tll(\mcV^*_2) \geq \dots \geq \tll(\mcV^*_S)$. Solving $\mcP$ faces a significant computational challenge since the cardinality of ${\mcF}$ grows exponentially with the system size $N$, the number of components $|\mcU|$, and the risk assessment horizon $P$.

\section{Sequential Search via $\pomdps$}
\label{sec:pomdp Model for FC Search}

\subsection{Sequential Search}
\label{sec:Sequential Search}

To circumvent the complexity of solving $\mcP$ in \eqref{eq:OBJ1}, we design an agent-based learning algorithm that sequentially constructs the set of $\fcs$ $\{\mcV^*_1,\dots,\mcV^*_S\}$. Specifically, the agent starts with constructing $\mcV^*_1\dff \langle \mcU_{1,1},\dots, \mcU_{1,P} \rangle$ such that it sequentially identifies the sets $\{\mcU_{1,1},\dots, \mcU_{1,P}\}$ in each stage $i\in[P]$ as follows. The agent admits the topology and the system state as its initial baseline inputs, denoted by $\mcG_0$ and $\bX_0$, respectively. In the first stage, the agent identifies~\emph{the components} in $\mcU$ removing which is expected to impose the most intense $\tll$. To control the complexity and reflect the reality of $\fcs$, in which failures occur component-by-component, we are interested in identifying only one component in each stage. Nevertheless, due to the physical constraints, removing one component can possibly cause outages in one or more other components in the same stage. We denote the set of all components to be removed in the first stage by the set $\mcU_{1,1}$.

The risk associated with each candidate set $\mcU_{1,1}$ has two, possibility opposing, impacts. The first pertains to the immediate loss due to component failures in $\mcU_{1,1}$, and the second captures the losses associated with the future possible failures driven by the failures in  $\mcU_{1,1}$. Hence, identifying the sets $\mcU_{1,1}$ involves look-ahead decision-making and cannot be carried out greedily based on only the immediate $\loss$s. Once the set $\mcU_{1,1}$ is identified {(via our proposed Algorithm~1 the details of which we discuss in Section~\ref{sec:Graph Recurrent Q learning})}, the agent removes all the components in this set to update the grid topology to $\mcG_1$, and uses simulations (by solving a power-flow or an optimal power-flow, if necessary) to determine the associated system state $\bX_1$. 

\begin{figure}[t]
 	\centering
    \includegraphics[width=0.6\linewidth]{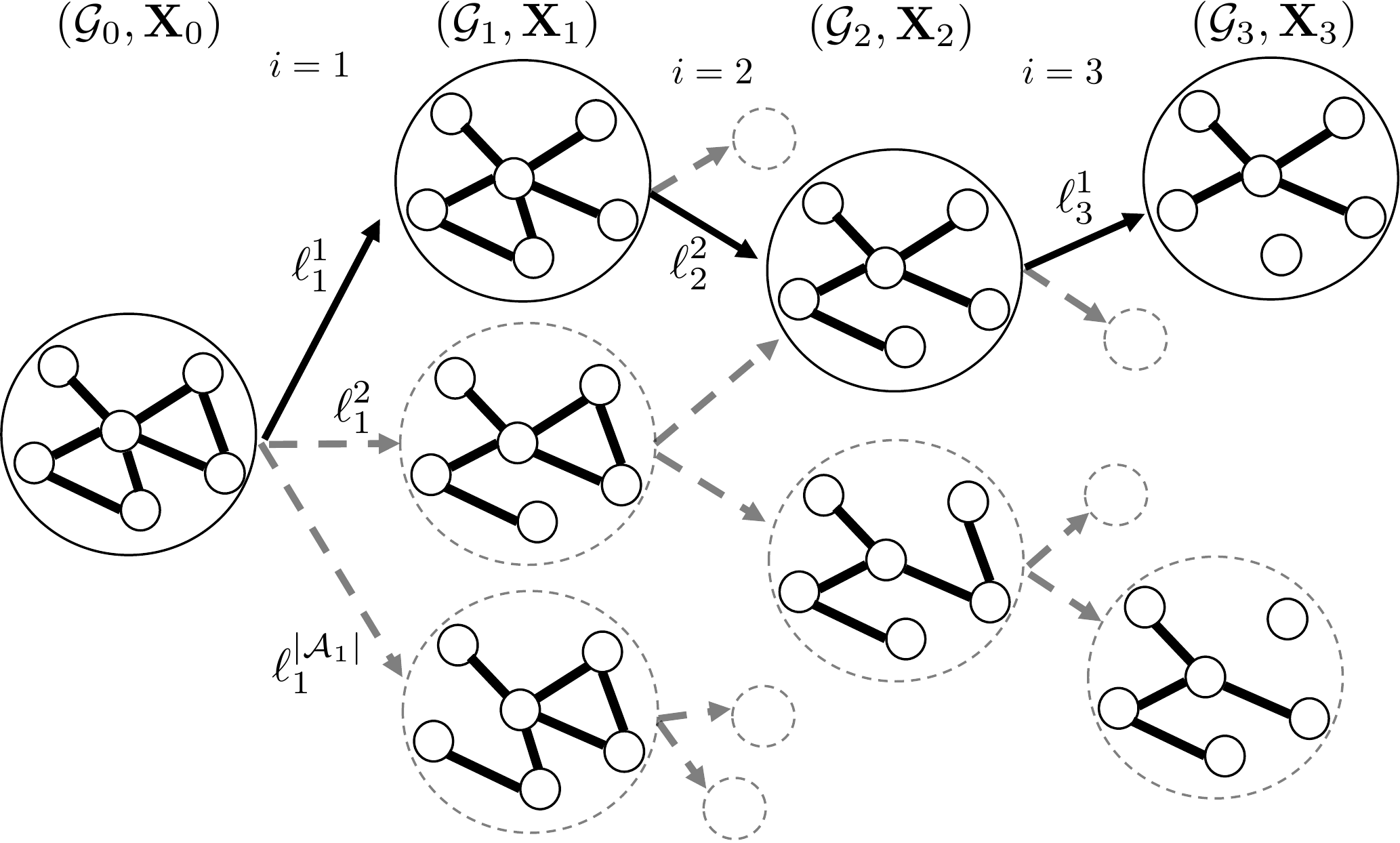}
    \caption{An agent decision process rendering a $\fc$ $\mcV^*_{s} = \langle \ell^{1}_{1}, \ell^{2}_{2}, \ell^{1}_{3}\rangle$.}
    \label{fig:look ahead search}
\end{figure}

Subsequently, $\mcG_1$ and $\bX_1$ are leveraged to identify the set $\mcU_{1,2}$ by removing~\emph{the components} from the set $\mcU \backslash \{\mcU_{1, 1}\}$, and this process continues recursively for a total of $P$ stages, at the end of which the set $\mcV^*_1$ is constructed. Subsequently, a similar process is repeated to construct $\mcV^*_2$ and the algorithm repeats this process $S$ times to identify $S$ sequences of interest. While repeating the search process, it is important that the agent avoids finding the same $\fc$ sequences over and over again that were discovered previously. Accordingly, as discussed in detail in Section~\ref{sec:Fault Chain Search Strategy}, we alter the agent decision process to take into account the number of times any given component was removed. Fig.~\ref{fig:look ahead search} illustrates a search process where an agent constructs a $\fc$ sequence $\mcV^*_{s} = \langle \ell^{1}_{1}, \ell^{2}_{2}, \ell^{1}_{3}\rangle$ by leveraging the current system state $(\mcG_{i}, \bX_{i})$ in each stage $i\in [3]$.


\subsection{Modeling Search as a POMDP}
\label{sec:Modeling Search as a POMDP}

The cascading outage process renders temporal dependencies across outage stages, typically spanning more than two stages. The full extent of such dependencies might be hidden in the observed (time) domain. We leverage the hidden dependencies in the latent (hidden) space of the fault chain generation process. Since the observation at each stage provides only~\emph{partial} information for decision making, we formalize the agent decision process at every stage of the search process by a partially observed Markov decision process ($\pomdp$).  In our search process, to control the computational complexity, we determine the system state in stage $(i+1)$, i.e., $(\mcG_{i+1}, \bX_{i+1})$ by leveraging the system state $(\mcG_{i}, \bX_{i})$ in stage~$i$. Nevertheless, the load loss at stage $(i+1)$ depends on all the past $i$ stages and the set of components removed in those~{system} states. In this subsection, we characterize the $\pomdp$ of interest, and address solving it in Section~\ref{sec:GRNN-based Approach to Solving POMDP}.

We denote the partial observation that the agent uses at stage~$i$ to determine the system state at stage $i+1$ by $\bO_i \dff (\mcG_i, \bX_i)$. Accordingly, we define the sequence
\begin{align}
    \label{eq:system state}
    \bS_{i} \dff \langle \bO_{0}, \dots, \bO_{i} \rangle\ ,
\end{align}
which we refer to as the $\pomdp$ state at stage $i$, and it characterizes the entire past sequence of observations that render $\bO_{i+1}$.  As stated earlier, at stage $i$ only $\bO_i$ is known to the agent. At stage $i$, upon receiving the observation $\bO_{i}$, the agent aims to choose a component from the set of~\emph{available} components to be removed in the next stage. To formalize this process, we define the agents action as its choice of the component of interest. We denote the action at stage $i$ by $a_i$. Accordingly, we define the action space $\mcA_{i}$ as the set of all remaining components, i.e., $\mcA_{i} \dff \mcU\backslash \{\cup_{j=1}^{i-1}\mcU_j\}$. Once the agent takes an action $a_{i} \in \mcA_{i}$ in stage $i$, the underlying $\pomdp$ state in next stage is randomly drawn from a transition probability distribution $\P$
\begin{align}
    \label{eq:system dynamics}
    \bS_{i+1} \sim \P(\bS \;|\;\bS_{i}, a_{i})\ .
\end{align} Probability distribution $\P$ captures the randomness due to the power system dynamics, and it is determined by the generator re-dispatch strategy in each stage of the $\fc$. To quantify the risk associated with taking action $a_i$ in $\pomdp$ state $\bS_{i}$ when transitioning to $\bS_{i+1}$, we define an instant reward $r_i$
\begin{align}
\label{eq:rewards}
r_i \dff  r(\bS_{i+1}\;|\; \bS_{i}, a_{i}) \dff \load(\mcG_{i}) - \load(\mcG_{i+1})\ .
\end{align} Hence, for any generic action selection strategy $\pi$, the aggregate reward collected by the agent starting from the baseline $\pomdp$ state can be characterized by a value function 
\begin{align}
    \label{eq:value function}
    V_{\pi}(\bS_{0}) \dff \sum_{i=0}^{P-1} \; \gamma^{i} \cdot r(\bS_{i+1}\;|\; \bS_{i}, \pi(\bO_{i}))\ ,
\end{align} where the discount factor $\gamma\in\R_+$ decides how much future rewards are favored over instant rewards, and $\pi(\bO_{i})$ denotes the action selected by the agent  given an observation $\bO_{i}$ in stage $i \in [P]$. Therefore, finding an optimal action selection strategy $\pi^{*}$ for the agent can be formally cast as solving
\begin{align}
    \label{eq:optimal policy}
    \mcQ: \quad \pi^{*} \dff \argmax_{\pi} \; \E \left[ V_{\pi}(\bS_{0}) \right]\ .
\end{align} 

\section{$\grnn$-based Approach to Solving $\pomdps$}
\label{sec:GRNN-based Approach to Solving POMDP}

\subsection{Motivation}
\label{sec:Motivation}

An optimal strategy $\pi^{*}$~\eqref{eq:optimal policy} in a $\pomdp$ environment can be found by leveraging classical dynamic programming algorithms. Such traditional solutions, however, not only require the knowledge of the transition probability model $\P$ and the reward function dynamics but also pose a significant computational challenge when solving for $\pi^{*}$. Under such unknown model settings, $Q$-learning serves as an effective~\emph{model-free value-based} $\rl$ algorithm~\cite{sutton:2018} that can find optimal strategies, although only in an $\mdp$ environment. In particular, the algorithm focuses on implicitly~\emph{estimating} the~{$\mdp$} state value function~\eqref{eq:value function}, from each state $\bS_{i}$, associated with every action $a_{i}\in \mcA_{i}$. These~\emph{estimates} are widely referred to as $Q$-values, denoted by $Q(\bS_{i}, a_{i}) \in \R$, where a higher $Q(\bS_{i}, a_{i})$ indicates that taking an action $a_{i}$ from a~{$\mdp$} state $\bS_{i}$ yields better aggregate future rewards~\cite{sutton:2018}.

Note that, however, when solving $\mcQ$ in a $\pomdp$ environment, the ordinary $Q$-learning algorithm is ineffective as partial observations $\bO_i$ are typically not reflective of the underlying $\pomdp$ state $\bS_i$. As a result, $Q(\bO_{i}, a_{i}) \neq Q(\bS_{i}, a_{i})$. Therefore, designing real-time $\fc$ search algorithms critically hinges on finding an~\emph{accurate} and~\emph{efficient} solution to $\mcQ$ in~\eqref{eq:optimal policy}. To this end, we aim to design an approach that judiciously exploits the underlying structure of each $\pomdp$ state $\bS_i$. In particular, we develop a graph recurrent $Q$-network $(\GRQN)$ architecture that exploits the following three key elements: 

\subsubsection{Neural Networks for $Q$-value Prediction} Implementing the ordinary $Q$-learning algorithm relies on building $Q$-tables that store $Q$-values for each~{$\pomdp$} state-action pair. However, storing such a $Q$-table imposes a significant storage challenge even for moderate system sizes since the number of state-action pairs scales combinatorially with the system size $N$. Furthermore, such an algorithm fails to incorporate continuous~{$\pomdp$} state spaces $\bS_{i}$. As a remedy, we employ universal function approximators such as $\nns$ to~\emph{predict} $Q(\bS_{i}, a_{i})$ values as approximate surrogates for each~\emph{estimated}~{$\pomdp$} state-action element in the $Q$-Table~\cite{Mnih:2015}.

\subsubsection{Spatial Correlation} The data streams generated by measurement units across the network have strong spatial correlation induced due to the inherent meshed topology of transmission networks. Specifically, the coordinates of $\bX_i$, in each stage $i$, are statistically correlated, and hence, data streams are typically jointly modeled via probabilistic graphical models~\cite{Miao:2011, Tajer:2018}. Therefore, it is important to judiciously leverage the spatial structure of $\bX_i$ when solving $\mcQ$.

\subsubsection{Temporal Correlation} Besides the spatial correlation, there exists a strong temporal structure induced due to the observational dependencies across the successive stages of any $\fc$ sequence. In particular, the load loss incurred due to a component failure in stage $i$ depends on the set of components that have failed in~\emph{all} the preceding $(i-1)$ stages.

To this end, in order to exploit the spatio-temporal correlation among the coordinates of a sequence of system states $\bX_{i}$ to~\emph{predict} $Q(\bS_{i}, a_{i})$ values accurately, we leverage the output of a time-varying $\grnn$ to incorporate the spatial structure and leverage its recurrency to effectively integrate observational dependencies through time to account for the temporal structure. Subsequently, the $\grnn$ output then acts as an input to a $\nn$ that~\emph{predicts} $Q$-values for each~{$\pomdp$} state-action pair, altogether assembling into an $\GRQN$ architecture. Next, we discuss the various components of the time-varying $\grnn$, and discuss designing a learning algorithm to search for $\fc$ sequences of interest in Section~\ref{sec:Graph Recurrent Q learning}.

\subsection{Development of a Time-Varying $\grnn$ Model}
\label{sec:Development of a grnn model}

$\grnns$ are a family of $\gnn$ architectures specialized for processing sequential graph structured data streams~\cite{Ruiz:2020}. These architectures exploit the~\emph{local} connectivity structure of the underlying network topology $\mcG_i$ to efficiently extract features from each bus by sequentially processing time-varying system states $\bX_i$ that evolve on a sequence of graphs $\mcG_{i}$. More broadly, these architectures generalize recurrent neural networks~\cite{recurrent:2001} to graphs. To lay the context for discussions, we first discuss time-varying graph convolutional neural networks $(\gcnns)$, the components of which serve as an essential building block to motivate time-varying $\grnns$.
\subsubsection{Time-Varying $\gcnns$} Consider the $i^{\rm th}$ stage of a $\fc$ sequence $\mcV_{s}$ where an agent receives an observation $\bO_i \dff (\mcG_i, \bX_i)$ on graph $\mcG_i$ associated with an adjacency matrix $\bB_{i}$. Central to the development of $\gcnns$ is the concept of a~\emph{graph-shift} operation that relates an input system state $\bX_{i}\in \R^{N\times F}$ to an output system state $\bF_{i} \in \R^{N\times F}$ 
\begin{align}
    \label{eq:graph shift}
    \bF_{i} \dff \bB_{i} \cdot \bX_{i}\ .
\end{align} Clearly, the output system state $\bF_{i}$ is a locally shifted version of the input system state $\bX_{i}$ since each element $[\bF_{i}]_{u,f}$, for any bus $u\in V_{i}$ and parameter~$f$, is a linear combination of input system states in its $1$-hop neighborhood. Local operations, such as~\eqref{eq:graph shift} capture the $1$-hop structural information from $\bX_{i}$ by~\emph{estimating} another system state $\bF_{i}$ and are important because of the strong spatial correlation that exists between $[\bX_{i}]_{u,f}$ and its network neighborhood determined by $\mcG_i$. Alternative transformations such as that employed in~\cite{Dwivedi:2022} quantify the merits of estimating system states that capture the spatial structure of $\bX_i$. In order to capture the structural information from a broader $K$-hop neighborhood instead,~\eqref{eq:graph shift} can be readily extended by defining a time-varying~\emph{graph convolutional filter} function $\sf{H}$ $: \R^{N \times F} \rightarrow \R^{N \times H}$ that operates on an input system state $\bX_i$ to~\emph{estimate} an output system state $\filter(\mcG_i, \bX_i\;;\;\mcH)$
\begin{align}
    \label{eq:graph filter}
    \filter(\mcG_i, \bX_i\;;\;\mcH) \dff \sum_{k=1}^{K} \; \left[\bB_{i}^{k-1} \cdot \bX_{i}\right] \cdot \bH_{k}\ ,
\end{align} 
where $H$ denotes the number of output features~\emph{estimated} on each bus and $\mcH \dff \{\bH_{k} \in \R^{F \times H}: k \in [K]\}$ denotes the set of filter coefficients parameterized by matrices $\bH_{k}$~\emph{learned} from simulations where each coordinate of $\bH_{k}$ suitably weighs the aggregated system state obtained after $k$ repeated $1$-hop graph-shift operations performed on $\bX_{i}$. Note that~\eqref{eq:graph filter} belongs to a broad family of~\emph{graph-time filters}~\cite{Isufi:2017} that are polynomials in time-varying adjacency matrices $\bB_{i}$. 
There exists many types of~\emph{graph-time filters}~\cite{Gama:2021}. We employ~\eqref{eq:graph filter} due to its simplicity. Nevertheless,~\eqref{eq:graph filter} only captures simple linear dependencies within $\bX_{i}$. To capture non-linear relationships within $\bX_{i}$, time-varying $\gcnns$~\emph{compose} multiple layers of graph-time filters~\eqref{eq:graph filter} and non-linearities such that the output system state of each $\gcnn$ layer is given by 
\begin{align}
    \label{eq:gcnns}
    \Phi(\mcG_i, \bX_{i}\;;\;\mcH) \dff \sigma\left(\;\filter(\mcG_i, \bX_{i}\;;\;\mcH)\;\right)\  ,
\end{align} where $\sigma:\R \rightarrow \R$ is the activation function (applied element-wise) such that $\Phi(\mcG_i, \bX_{i}\;;\;\mcH) \in \R^{N\times H}$.

\subsubsection{Time-Varying $\grnns$} $\gcnns$ can only extract spatial features from each system state $\bX_{i}$ independently~\eqref{eq:gcnns}. For this reason, we add recurrency to our $\gcnn$ model~\eqref{eq:gcnns} in order to capture the temporal observational dependencies across the various stages of a $\fc$ sequence to construct a $\grnn$. A time-varying $\grnn$ extracts temporal features from an input sequence $\langle \bO_i:i \in [P]\rangle$ by~\emph{estimating} a sequence of hidden system states $\langle\bZ_{i}:i \in [P]\rangle$ where each~{system} state $\bZ_{i}\in \R^{N\times H}$ is latent that facilitates in summarizing the entire past observational history, that is both redundant and difficult to store, until stage $i$. This is done by judiciously parameterizing each hidden system state $\bZ_{i}$ that is a function of the graph-time filter output~\eqref{eq:graph filter} operated on both the current input system state $\bX_i$ and previous hidden system state $\bZ_{i-1}$ independently to obtain
\begin{align}
    \label{eq:GRNN time varying hidden state}
    \bZ_{i} \dff \sigma\left(\;\filter_{1}(\mcG_i, \bX_{i}\;;\;\mcH_{1}) \; + \; \filter_{2}(\mcG_{i-1}, \bZ_{i-1}\;;\;\mcH_{2})\;\right)\ ,
\end{align} where $\filter_{1}:\R^{N\times F} \rightarrow \R^{N\times H}$ and $\filter_{2}:\R^{N\times H} \rightarrow \R^{N \times H}$ are filters~\eqref{eq:graph filter} each parameterized by a distinct set of filter coefficients $\mcH_{1}=\{\bH^{1}_{k} \in \R^{F \times H}\; \forall\;k \in [K]\}$ and $\mcH_{2}=\{\bH^{2}_{k} \in \R^{H \times H}\; \forall\;k \in [K]\}$, respectively. Subsequently, similar to~\eqref{eq:gcnns}, to capture the non-linear relationships from each hidden system state $\bZ_{i}$ to facilitate dynamic decision-making on graphs $\mcG_i$, the~\emph{estimated} output system state $\bY_{i}\in \R^{N \times G}$
\begin{align}
    \label{eq:GRNN output}
    \bY_{i} \dff \rho\left(\; \filter_{3}(\mcG_i, \bZ_{i}\;;\; \mcH_{3})\;\right) \quad \forall i \in [P]\ ,
\end{align} where the graph-time filter $\filter_{3}:\R^{N\times H} \rightarrow \R^{N \times G}$ in~\eqref{eq:graph filter} is parameterized by the filter coefficient set $\mcH_{3}$, $G$ denotes the number of~\emph{estimated} output features on each bus $u\in V_{i}$, and $\rho$ is a pointwise non-linearity applied element-wise to output $\filter_{3}$. Note that $H$, $K$, $G$, $\sigma$ and $\rho$ are hyper-parameters for a $\grnn$. Additionally, the number of~\emph{learnable} parameters $\mcH_{i} \; \forall i \in [3]$ is independent of the system size $N$ and the horizon $P$ of the $\fc$ sequence due to parameter sharing across the stages of the $\fc$, providing the model with flexibility to~\emph{learn} from input sequences $\langle \bO_i:i \in [P]\rangle$ of different and long risk assessment horizons without a combinatorial growth in the number of~\emph{learnable} parameters, ensuring tractability. Next, we leverage the sequence of $\grnn$ output system states $\langle \bY_{i}:i\in [P]\rangle$ to learn a strategy $\tilde \pi \approx \pi^{*}$~\eqref{eq:optimal policy} to efficiently solve~\eqref{eq:OBJ1}.

\subsection{Finding a Strategy via Graph Recurrent $Q$-learning}
\label{sec:Graph Recurrent Q learning}

As discussed in Section~\ref{sec:GRNN-based Approach to Solving POMDP}, estimating $Q(\bS_i, a_i)$ via traditional $\rl$ algorithms is intractable. Hence, we choose to~\emph{predict} $Q(\bS_i, a_i)$ via $\nns$. However, since the agent is no longer privy to the true underlying $\pomdp$ state $\bS_{i}$, inspired by the approach of~\cite{DRQN:2015}, we develop a $\GRQN$ architecture to~\emph{estimate} $\bY_{i}$~\eqref{eq:GRNN output} in order to~\emph{predict} $Q(\bY_{i}, a_{i})$ as a proxy to approximate $Q(\bS_{i}, a_{i})$. Additionally, since for any given initial system state $(\mcG_0, \bX_0)$, it is of interest to find $S$ {number of} $\fc$ sequences~\eqref{eq:OBJ1} with the maximum $\tll$, it is not enough to~\emph{predict} $Q(\bY_{i}, a_{i})$ but rather to guide the search of subsequent $\fcs$ with the next highest $\tll$ by leveraging $Q(\bY_{i}, a_{i})$. Therefore, we design a graph recurrent $Q$-learning algorithm to discover the $S$ {number of} $\fc$ sequences of interest $\{\mcV^*_1,\dots,\mcV^*_S\}$ one after the other while concurrently training the $\GRQN$. Next, we discuss the $\GRQN$ architecture and the designed training algorithm. 

\begin{figure}[t]
 	\centering
    \includegraphics[width=.8\linewidth]{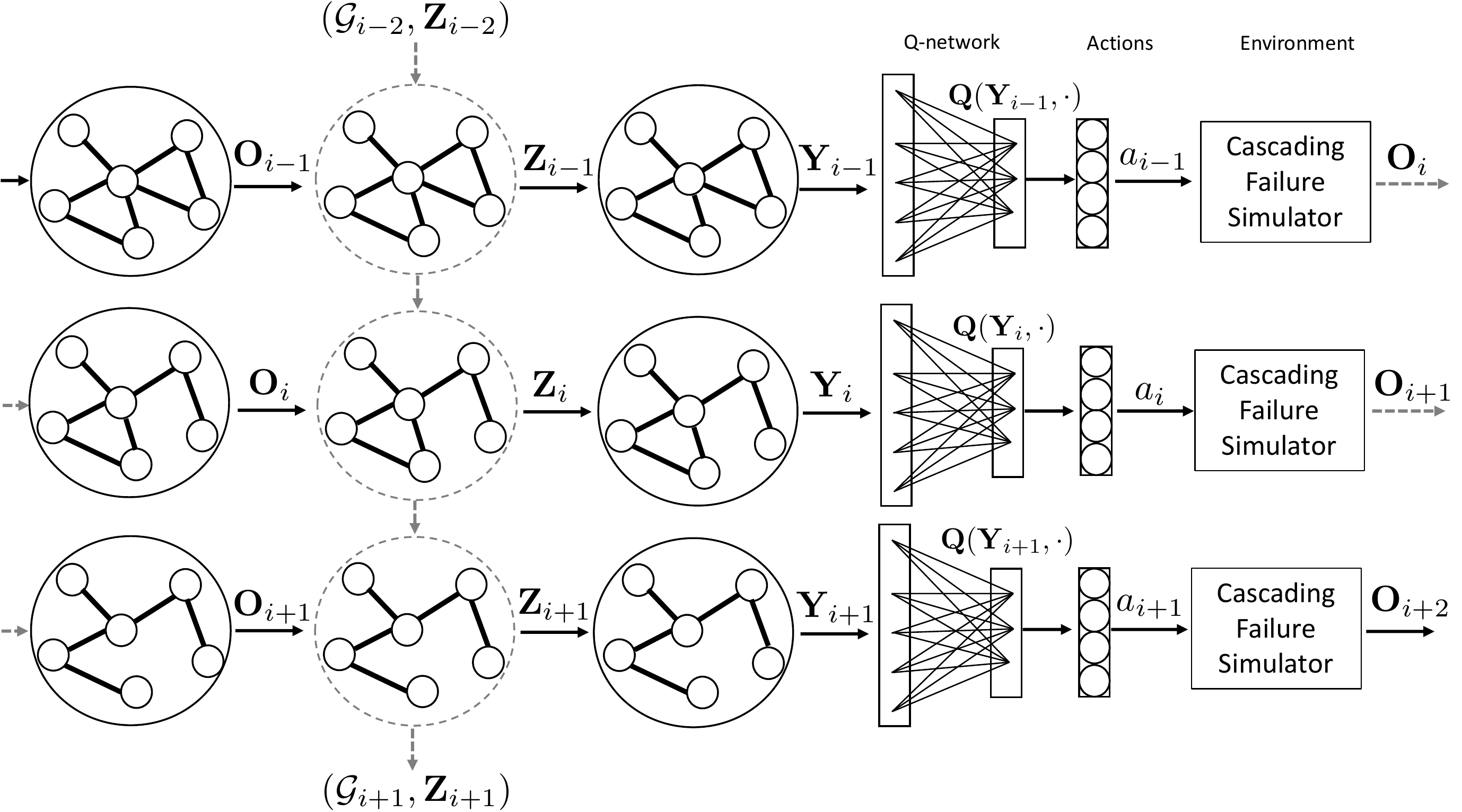}
    \caption{Graph recurrent $Q$-network $(\GRQN)$ architecture.}
    \label{fig:GRQN arch}
\end{figure}

\subsubsection{$\GRQN$ Architecture} The architecture consists of a time-varying $\grnn$, parameterized by $\btheta_{\grnn} \dff \{\mcH_{i} : i \in [3]\}$, that sequentially processes observations at each stage of the $\fc$ and a fully-connected $\nn$, parameterized by $\btheta_{\nn}$, to~\emph{predict} $Q(\bY_{i}, a_{i})$ for each action $a_{i}\in \mcA_{i}$. Accordingly, we parameterize the $\GRQN$ by $\btheta \dff \{\btheta_{\grnn}, \btheta_{\nn}\}$. For any $\fc$ sequence $\mcV_{s}$, an input data stream of observations $\bO_{i}\in \R^{N \times F}$ obtained at each stage $i$ of the $\fc$ acts as an input to the $\grnn$ using which an output system state $\bY_{i} \in \R^{N \times G}$ is estimated. Subsequently, a fully connected $\nn$ of input and output dimension $N \times G$ and $|\mcU|$, respectively, is leveraged to output a $\bQ(\bY_{i}, \cdot | \btheta) \in \R^{|\mcU|}$ vector consisting of $Q$-values for each action. Overall, at each stage $i$, a $\GRQN$ takes $\bO_{i}$ and $\bZ_{i-1}$ as it's baseline inputs, concisely denoted by $\GRQN(\bO_{i}, \bZ_{i-1}| \btheta)$, and outputs a vector $\bQ(\bY_{i}, \cdot|\btheta)$ and the next hidden~{system} state $\bZ_{i}$~\eqref{eq:GRNN time varying hidden state}, crucial to carry forward to guide the search of $\fcs$ during the training of the $\GRQN$. Note that we initialize $\bZ_{0}$ by $\mathbf{0} \in \R^{N \times H}$. Fig.~\ref{fig:GRQN arch} illustrates the end-to-end $\GRQN$ architecture. 

\subsubsection{Sequential Experience Buffer} In order to deal with the issue of~\emph{catastrophic forgetting}~\cite{Mnih:2015}, we employ a~\emph{sequential} experience buffer that stores $\fc$ sequences discovered during training of the $\GRQN$ from which batches of random sequences are sampled to facilitate the learning of parameters $\btheta$. While there exists various ways to implement such a buffer, we employ an ordered list that stores all the visited transition tuples as a sequence $\langle (\bO_{i}, a_{i}, r_{i}, \bO_{i+1}, \isend(\bO_{i+1})):i \in [P]\rangle$ where we have defined $\isend(\bO_{i+1})$ as a boolean value, if true, indicating that the observation $\bO_{i+1}$ is associated with a last stage of the risk assessment horizon $P$.

\subsubsection{Training the $\GRQN$} For stability in the training of the $\GRQN$, we employ the standard trick~\cite{Mnih:2015} of splitting the task of predicting and evaluating $Q$-values via two separate $\GRQNs$, a target network $\GRQN(\cdot, \cdot|\btheta^{-})$ and a behavior network $\GRQN(\cdot, \cdot|\btheta)$ each parameterized by a distinct set of parameters $\btheta^{-}$ and $\btheta$, respectively. For every training iteration $n$ of the graph recurrent $Q$-learning algorithm, the agent samples $B$ random batches of $\fc$ sequences, each of type $\langle (\bO_{j}, a_{j}, r_{j}, \bO_{j+1}, \isend(\bO_{j+1})):j \in [P]\rangle$, from the sequential experience buffer on which the target $\GRQN$ is~\emph{unrolled} to estimate $\bY_j$~\eqref{eq:GRNN output} using which $B$ batches of $\bQ(\bY_j, \cdot|\btheta^{-})$ are predicted with respect to the target network. Subsequently, the agent computes a~\emph{look-ahead} target output for each batch where each target $t_{j} \; \forall j \in [P]$ is given by
\begin{align}
    \label{eq:target update}
    t_{j} = r_{j} +  \gamma \cdot (1 - \isend(\bO_{j+1})) \cdot \max_{a}Q(\bY_{j+1}, a | \btheta^{-})\ .
\end{align} Accordingly, the parameters of the behavior  network is updated via gradient descent with respect to a quadratic loss
\begin{align}
    \label{eq:gradient update}
    \btheta_{n+1} & = \btheta_{n} - \alpha \cdot \nabla_{\btheta_{}}(t_{j} - Q(\bY_{j}, a_{j} | \btheta_{}))^{2}\ , 
\end{align} where $\alpha$ denotes the learning rate and $n$ denotes the current training iteration. The update~\eqref{eq:gradient update} is preformed $\kappa$ times for every action taken by the agent in any stage $i\in[P]$ of the $\fc$ and additionally, serves as a means to control the~\emph{computational complexity} of our learning algorithm. In this paper, we employ the Adam optimizer~\cite{adam:2015} to perform the gradient update~\eqref{eq:gradient update} and update the target network parameters $\btheta^{-} = \btheta$ at the end of every $\fc$ sequence discovered. 

\setlength{\textfloatsep}{0pt}
\begin{algorithm}[t]
	\caption{Graph Recurrent $Q$-learning ($\GRQN$)}
	\begin{algorithmic}[1]
		\footnotesize
		\Procedure{Graph Recurrent $Q$-learning}{} 
			\State Initialize behaviour  network with random $\btheta$ $\GRQN(\cdot, \cdot | \btheta)$
			\State Initialize target network with weights $\btheta^{-} = \btheta$ $\GRQN(\cdot, \cdot | \btheta^{-})$ 	
            \State Initialize $\sf{Buffer} \leftarrow \langle \rangle$ 	
            \State Initialize $\bZ_{0} \leftarrow \mathbf{0} \in \R^{N\times H}$ 	
            \For {$\episode \; s = 1, \dots, S$} 
		        \State $\episode \leftarrow \langle \rangle$  
		        \State $\mcV_{s} \leftarrow \langle \rangle$
                \State Reset power-flow according to initial state $\bS_{0}, \bO_{0}$
    			\For {Stage $i = 1, \dots, P$} 
    			    \State $\_\_, \bZ_{i} \leftarrow \GRQN(\bO_{i}, \bZ_{i-1} | \btheta)$
				    \State $a_{i} \leftarrow 
                    \begin{cases}
                        \mbox{explore} & \text{if rand}(0, 1) \leq \epsilon \\
                        \mbox{exploit} & \text{otherwise}
                    \end{cases}$
 	                \State Take action $a_{i}$ and determine $\mcU_{i}$
 	                \State $\mcV_{s} \leftarrow  \mcV_{s} \cup \mcU_{i}$
            	    \State Update power-flow and obtain $\bO_{i+1}$
            	    \State Calculate load loss $r_{i}$ from~\eqref{eq:rewards}
            	    \State $\episode \leftarrow \episode \cup (\bO_{i}, a_{i}, r_{i}, \bO_{i+1}, \isend(\bO_{i+1}))$
            		\If{$s \geq {\sf Explore}$}
            		    \State ${\sf count}(\bS_{i},a_{i}) \leftarrow {\sf count}(\bS_{i},a_{i}) + 1$ 
                		\For {training $n = 1, \dots, \kappa$} 
                			\State Sample $B$ $\fc$ sequences from $\buffer$
                			\State $t_{j}, \_\_ \leftarrow \GRQN(\bO_{j+1}, \mathbf{0} | \btheta^{-})$ from~\eqref{eq:target update}
                            \State $Q(\bY_{j}, a_{j}| \btheta), \_\_  \leftarrow \GRQN(\bO_{j}, \mathbf{0} | \btheta)$
                            \State Calculate $\nabla_{\btheta_{}}(t_{j} - Q(\bY_{j}, a_{j} | \btheta))^{2}$ 
                            \State Update $\btheta$ as in~\eqref{eq:gradient update}
                            \State Update $\epsilon$ as in~\eqref{eq:epsilon update} 
                    	\EndFor	
            		\EndIf
	        	    \State Update availability of actions backwards
            	\EndFor	
            	\State $\sf{Buffer} \leftarrow \sf{Buffer} \cup \episode$
            	\State $\bZ_{0} \leftarrow \bZ_{P}$
        		\If{$\tll(\mcV_{s}) \geq M$}
        		    \State Store risky $\fc$
        		\EndIf
        		\State $\btheta^{-} = \btheta$
            \EndFor
		\EndProcedure
		\State Find the accumulated risk due to all the $S$ $\fcs$ $\{\mcV_1,\dots,\mcV_S\}$.
	\end{algorithmic}
\end{algorithm}
\subsubsection{Graph Recurrent Hidden~{System} State Updates} As the agent gains more experience and continues to store visited transition sequences of tuples in the experience buffer, the hidden system state $\bZ_{i}$ of the $\grnn$ may either zeroed or carried forward after every newly discovered $\fc$. Our experiments suggest that sequential updates where the hidden~{system} state $\bZ_{i}$~\eqref{eq:GRNN time varying hidden state} is carried forward from the previous stages throughout the parameter update~\eqref{eq:gradient update} leads to learning of better $\fc$ search strategies. Hence, we choose to carry forward the previously learned hidden system state during the training of our $\GRQN$.

\subsubsection{{Outline of Algorithm~1}}

Algorithm~1 outlines the graph recurrent $Q$-learning algorithm used to~\emph{learn} the parameters $\btheta$ of the behaviour $\GRQN(\cdot, \cdot | \btheta)$ to facilitate the discovery of the $S$ {number of} $\fc$ sequences of interest. During the initial few iterations, since the sequential experience buffer is empty, we allow the agent to~\emph{explore} and fill the buffer with $\fc$ sequences for ${\sf Explore}$ number of iterations~\emph{offline} (more details in Section~\ref{sec:Offline Sequential Buffer Initialization}). Subsequently, the real-time algorithm is initiated. {Initially, the agent lacks any information about the cascading failure dynamics. Hence, it relies on the~\emph{prior knowledge} to construct~$\mcU_{j,i}$, for any fault chain $j$ in each stage $i\in [P]$, as the agent is unable to evaluate the $\loss$ associated with removing an arbitrary component $\ell_{i} \in \mcU_j \backslash \{\cup_{k=1}^{i-1}\mcU_{j, k}\}$ in any stage of the $\fc$ $\mcV_{j}$. Gradually, Algorithm~1 learns to construct the sets~$\mcU_{j,i}$ by leveraging the learned latent graphical feature representations. These representations are obtained by optimizing the~\emph{look-ahead} target function~\eqref{eq:target update} characterized by the behavior network $\GRQN(\cdot, \cdot|\btheta)$ since the parameters $\btheta$ of this network determine the $Q$-values influencing the actions $a_i \in \mcA_i$ taken by the agent. Therefore, when choosing actions $a_i \in \mcA_i$, the agent should make a trade-off between~\emph{exploration} and~\emph{exploitation} throughout the training of the $\GRQNs$.} Next, we discuss an~\emph{exploration-exploitation} search strategy that the agent employs to make the real-time search of $\fcs$ of interest more efficient. 


\subsection{Fault Chain Search Strategy}
\label{sec:Fault Chain Search Strategy}

We employ the standard $\epsilon$-greedy search strategy with an adaptive exploration schedule. Initially, the agent is compelled to take actions based on prior knowledge to find $\fcs$ with maximum expected $\tll$. Typically, since an outage of a component carrying higher power makes the remaining components vulnerable to overloading, a reasonable exploration strategy of the agent would be to remove components carrying maximum power-flow. Therefore, we follow a power-flow weighted $(\PFW)$ exploration strategy (also adopted in~\cite{Zhang:2020}) such that, in any stage $i$ of the $\fc$, the agent chooses the $j^{\rm th}$ available component $\ell^{j}_{i} \in \mcA_{i}$ according to the rule
\begin{align}
    \label{eq:exploration policy}
    a_{i} = \argmax_{\ell^{j}_{i}} \; \frac{ \pf(\ell^{j}_{i})/\sqrt{{\sf count}(\bS_{i},\ell^{j}_{i}) + 1} }{\sum_{k=1}^{|\mcA_{i}|} \; \pf(\ell^{k}_{i})/\sqrt{{\sf count}(\bS_{i},\ell^{k}_{i}) + 1} }\ ,
\end{align} with probability $\epsilon$ where we denote $\pf(\ell^{j}_{i})$ as the~\emph{absolute} value of the power flowing through component $\ell^{j}_{i}$ and denote ${\sf count}(\bS_{i},\ell^{j}_{i})$ as the number of times the component $\ell^{j}_{i}$ was chosen when the agent was in~{$\pomdp$} state $\bS_{i}$ in the past. On the other hand, as the agent gains more experience, the agent should choose actions based on the $Q$-values learned via the behaviour network $\GRQN(\cdot, \cdot | \btheta)$. Accordingly, a strategy based on $Q$-values learned by the agent is designed. Specifically, actions are chosen proportional to the $Q$-values normalized by each~{$\pomdp$} state-action visit count to~\emph{avoid repetitions} of $\fc$ sequences discovered earlier. Accordingly, the agent chooses action $a_{i}$
\begin{align}
    \label{eq:exploitation policy}
    a_{i} = \argmax_{\ell^{j}_{i}} \; \frac{Q(\bY_{i}, \ell^{j}_{i} | \btheta)}{\sqrt{{\sf count}(\bS_{i}, \ell^{j}_{i}) + 1}}\ ,
\end{align} with probability $1 - \epsilon$.

In order to balance the exploration-exploitation trade-off between~\eqref{eq:exploration policy} and~\eqref{eq:exploitation policy} during the training of the $\GRQN$, it is important to dynamically alter the probability $\epsilon$ so that, with more experience, the agent chooses actions based on~\eqref{eq:exploitation policy}. Appropriately, we follow the exploration schedule given by 
\begin{align}
    \label{eq:epsilon update}
    \epsilon = \max\left(\frac{\sum_{j=1}^{|\mcA_{1}|} \; \pf(\ell^{j}_{1})/\sqrt{{\sf count}(\bS_{0},\ell^{j}_{1}) + 1}}{\sum_{k=1}^{|\mcA_{1}|} \; \pf(\ell^{k}_{1})},\; \epsilon_{0}\right)\ , 
\end{align} where $\epsilon_{0}$ ensures a minimum level of exploration. 
{\small
\begin{table*}[t]
\centering
\scalebox{0.7}{
	\begin{tabular}{| c | c | c | c | c |} 			
		\hline
		& \multicolumn{2}{c|}{ \thead{Evaluation Metrics} } 
		& \multicolumn{2}{c|}{ \thead{Accuracy Metrics} }\\ [2.0ex]
		\hline
		\thead{Algorithm}               
		& \thead{Range for Accumulative $\tll$\\ $\sum_{i=1}^{S} \tll(\mcV_{i})$ (in $\mws$)}   
		&  \thead{Range for the No. of Risky\\ $\fcs$ $\sum_{i=1}^{S}\mathds{1}( \tll(\mcV_{i}) \geq M )$}  
		& \thead{Range for} ${\sf Regret}(S)$ (in $\mws$)
		& \thead{Range for} ${\sf Precision}(S)$ \\ [2.0ex] 
		\hline
		
		Algorithm~1$\;(\kappa = 3)$    & $110.42\times 10^{3}\;\pm\;37\%$  & $199\;\pm\;71$    & $765.33\times 10^{3}\;\pm\;5.3\%$   & $0.169\;\pm\;35\%$   \\ [1.0ex] 
		\hline
		
		Algorithm~1$\;(\kappa = 2)$    & $96.18\times 10^{3}\;\pm\;38\%$  & $173\;\pm\;59$  & $779.57\times 10^{3}\;\pm\;4.7\%$   & $0.144\;\pm\;34\%$   \\ [1.0ex] 
		\hline
		
		Algorithm~1$\;(\kappa = 1)$    & $86.43\times 10^{3}\;\pm\;35\%$  & $161\;\pm\;48$  & $789.32\times 10^{3}\;\pm\;3.87\%$   & $0.134\;\pm\;29\%$   \\ [1.0ex] 
		\hline
	    
	    $\PFW$ + $\rl$ + $\TE$ \cite{Zhang:2020} & $60.63\times 10^{3}\;\pm\;3.4\%$     & $109\;\pm\;5$ & $815.12\times 10^{3}\;\pm\;0.26\%$  & $0.0909\;\pm\;4.8\%$   \\[1.0ex] 
		\hline
		
		$\PFW$ + $\rl$ \cite{Zhang:2020}  & $57.18\times 10^{3}\;\pm\;3.1\%$    & $101\;\pm\;9$  & $818.57\times 10^{3}\;\pm\;0.22\%$ & $0.084\;\pm\;3.7\%$   \\ [1.0ex] 
		\hline
	\end{tabular}%
	}
	\caption{{Performance comparison for the IEEE-$39$ New England test system.}}
	\label{table:39 bus}
\end{table*}
}

\section{Case Studies and Discussion}
\label{sec:Case Studies and Discussion}

In this section, extensive simulations are performed to validate the proposed graphical framework to find $\fcs$ that incur large $\tll$s. While we employ the $\dc$ power-flow model to simulate $\fcs$, other models such as the $\ac$ power-flow can be easily integrated within our framework. In order to generate $\fcs$, we employ PYPOWER a port of MATPOWER~\cite{Zimmerman:2011} to  Python and leverage PyTorch~\cite{pytorch:2019} to train the behavior and target $\GRQNs$.

\subsection{{Algorithm Initialization and Evaluation Criteria}}
\label{sec:Algorithm Initialization and Evaluation Criteria}

\subsubsection{Offline Sequential Buffer Initialization}
\label{sec:Offline Sequential Buffer Initialization}

To initiate the parameter update of the behavior $\GRQN$ via gradient descent~\eqref{eq:gradient update}, there must exist at least $B$ $\fc$ sequences in the experience buffer. However, unlike in the case of~\eqref{eq:exploration policy}, the buffer can be populated~\emph{offline} for any loading condition. Therefore, the agent can afford to take actions greedily with respect to components conducting maximum power and, accordingly, backtrack to update the availability of actions to avoid repeating $\fcs$ discovered previously. Hence, prior to the start of our real-time $\fc$ search Algorithm~1, we let the agent explore~\emph{offline} for ${\sf Explore}$ iterations where the agent, in any stage $i$, chooses actions according to the rule
\begin{align}
    \label{eq:offline strategy}
    a_{i} = \argmax_{\ell^{j}_{i}} \; \frac{\pf(\ell^{j}_{i})}{\sum_{k=1}^{|\mcA_{i}|} \; \pf(\ell^{k}_{i})}\ ,
\end{align} with probability $1$ to fill the sequential experience buffer. Note that this needs to be done only once offline for any loading condition. This is important since the quality of the sequences in the buffer greatly affects the efficiency of the search.

\subsubsection{{Accuracy Metrics and Evaluation Criteria}}
\label{sec:Accuracy Metrics and Evaluation Criteria}

{We aim to identify $\fcs$ with the largest $\tll$s. Hence, one natural evaluation metric is the accumulated $\tll$ due to the set of $\fcs$ $\{\mcV_1,\dots,\mcV_S\}$.} Besides that, we also consider the total number of~\emph{risky} $\fcs$ discovered as a function of $\fc$ sequence iterations $s\in [S]$, as considered in~\cite{Zhang:2020}, to further perform comparisons. A $\fc$ sequence $\mcV_{s}$ is deemed~\emph{risky} if it's associated $\tll$ exceeds a pre-specified level $M$, i.e., $\tll(\mcV_{s})\geq M$. {These two metrics are useful for evaluating the relative performance of Algorithm~1 compared to alternative approaches.} {For evaluating the accuracy of Algorithm~1, we adopt the following two metrics that quantify the accuracy in load loss and risky fault chain discovery rates.}
{\paragraph{Load Loss Accuracy} We define a regret term that quantifies the gap between the accumulated $\tll$ discovered by Algorithm~1 and the~\emph{optimal} accumulated $\tll$s of the ground truth $\fcs$ with the maximum $\tll$ given by the set $\{\mcV^*_1,\dots,\mcV^*_S\}$. Specifically, for $s \in [S]$ we define
\begin{align}
    \label{eq:accuracy-1}
     {\sf Regret}(s) \dff \sum_{i=1}^{S} \tll(\mcV^{*}_{i}) - \sum_{i=1}^{s} \tll(\mcV_{i}) \ .
\end{align} A lower regret value indicates a higher accuracy. 
\paragraph{Risky $\fc$ Discovery Rate} We define a precision metric that quantifies the fraction of $\fcs$ that are deemed risky in the set of discovered $\fc$. Specifically, for $s \in [S]$ we define
\begin{align}
    \label{eq:accuracy-2}
    {\sf Precision}(s) & \dff \frac{1}{s} \sum_{i=1}^{s}\mathds{1}( \tll(\mcV_{i}) \geq M )\ ,
\end{align}
where $\mathds{1}(\cdot)$ denotes the indicator function. A higher precision value indicates higher accuracy.} 

\subsection{{IEEE-39 New England Test System}}
\label{sec:IEEE-39 New England Test System}

This test system comprises of $N = 39$ buses and $|\mcU| = 46$ components, including $12$ transformers and $34$ lines. We consider a loading condition of $0.55\times{\sf base\_load}$, where ${\sf base\_load}$ denotes the standard load data for the New England test case in PYPOWER~\emph{after} generation-load balance to quantify the performance of our approach. These loading conditions were chosen since it is relatively difficult to discover $\fcs$ with large $\tll$s in a lightly loaded power system as there are fewer such $\fcs$ in comparison to the space of all $\fc$ sequences $|{\mcF}|$.

\subsubsection{Parameters and Hyper-parameters} 
\label{sec:Parameters and Hyper-parameters}

The hyper-parameters of the $\GRQNs$ are chosen by performing hyper-parameter tuning. Accordingly, we choose $H = G = 12$ hidden and output number of features when computing both the hidden system state~\eqref{eq:GRNN time varying hidden state} and the output system state~\eqref{eq:GRNN output}. We use $K = 3$ graph-shift operations for the graph-filter~\eqref{eq:graph filter} that is used to compute~\eqref{eq:GRNN time varying hidden state} and~\eqref{eq:GRNN output}. We use both $\rho$ and $\sigma$ as the hyperbolic tangent non-linearity $\sigma = \rho = \tanh$ and the ${\sf ReLU}$ non-linearity for the fully-connected $\nn$ that approximates the $Q$-values. For other parameters, we choose $F=1$ since we employ voltage phase angles as the only input system state parameter $f$, choose $\gamma = 0.99$ since large $\loss$s mostly occur in the last few stages of the $\fc$ sequence, an $\epsilon_{0} = 0.01$ to ensure a minimum level of exploration during the $\fc$ search process, a batch size $B = 32$, ${\sf Explore} = 250$, a risk assessment horizon $P=3$, $\fc$ sequence iteration $S = 1200$ (\emph{excluding} the initial ${\sf Explore}$ iterations), learning rate $\alpha = 0.005$, and $\kappa \in [3]$ that controls the frequency of the learning update~\eqref{eq:gradient update} and also governs the computational complexity of the graph recurrent $Q$-learning algorithm. 

We consider $M$ equal to $5\%$ of total load (where the total load is $0.55\times{\sf base\_load}$).~{To quantify the regret~\eqref{eq:accuracy-1}, we need to compute the $\tll$ associated with the $S$ most critical $\fc$ sequences (ground truth) $\mcV^{*}_{s}$, $\forall s \in [S]$. This is carried out by generating~\emph{all possible} $\fc$ sequences (i.e., set $\mcF$) with a target horizon of $P=3$ for the considered total load of $0.55\times{\sf base\_load}$. By leveraging the pre-computed set $\mcF$, we observe a total of $3738$ risky $\fcs$ for the loading condition $0.55\times{\sf base\_load}$ using our developed $\fc$ simulator.}

\subsubsection{{Accuracy and Efficiency - Performance Results}} 
\label{sec:Accuracy and Efficiency - Performance Results}

\begin{figure}[t]
 	\centering
 	\includegraphics[width=0.65\linewidth]{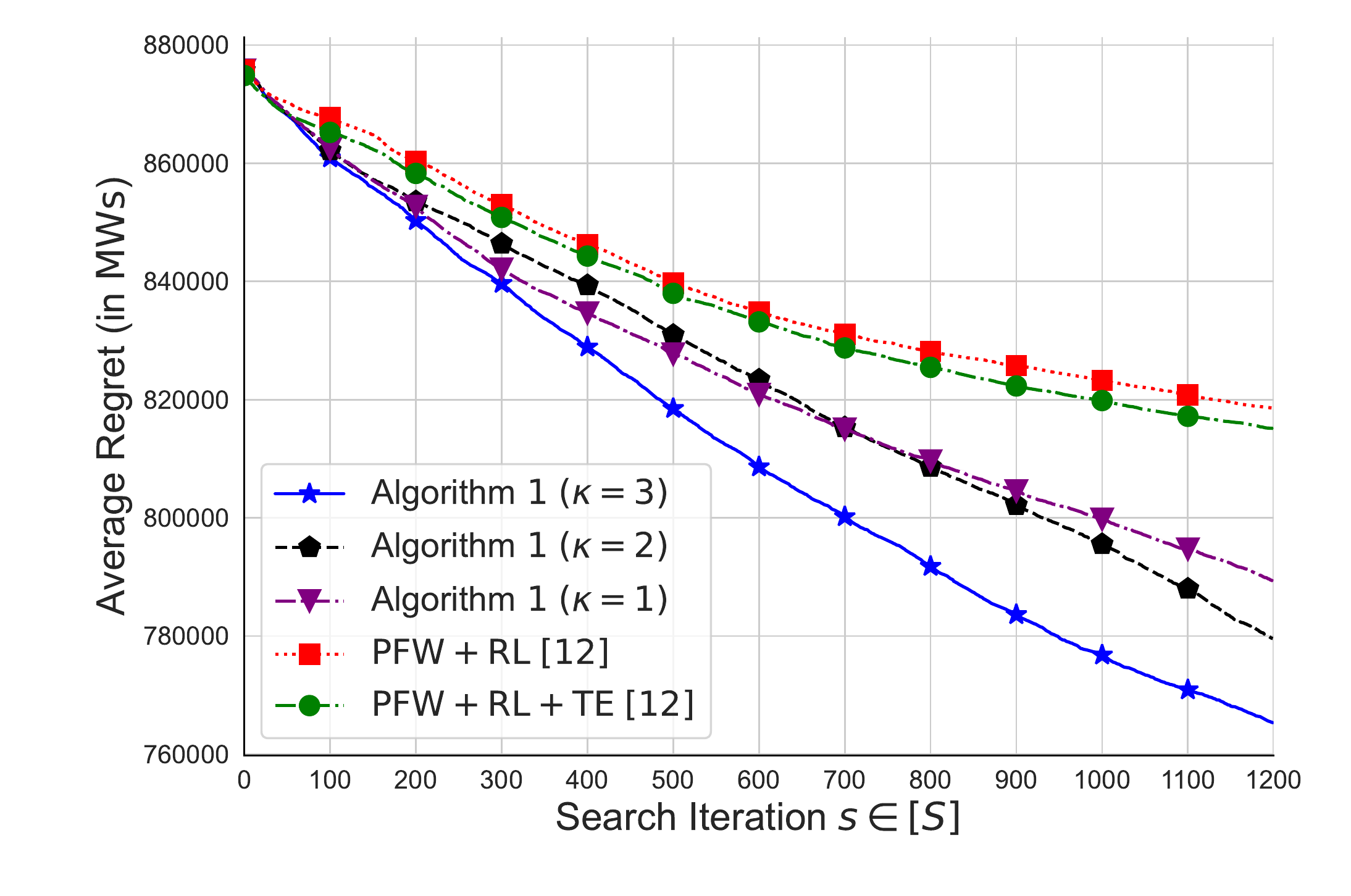}
 	\caption{{${\sf Regret}(s)$ versus $s$ for the IEEE-$39$ bus system.}}
 	\label{fig:Accum Risk}
 	\centering
\end{figure}

Prior to the start of Algorithm~1, we let the agent fill the experience buffer for ${\sf Explore}$ iterations following the strategy~\eqref{eq:offline strategy} and subsequently, initiate Algorithm~1. Table~\ref{table:39 bus} illustrates the results obtained. The first column specifies the algorithm employed for evaluation. {The second and third columns show the mean and standard deviation of the evaluation metrics and the forth and fifth columns show the mean and standard deviation of the accuracy metrics defined in~\eqref{eq:accuracy-1} and \eqref{eq:accuracy-2} for $S=1200$.} It is observed that Algorithm~1 with a greater $\kappa$ discovers $\fc$ sequences that incur larger accumulated $\tll$ and also discovers more number of risky $\fcs$, on average.~{This indicates that the accuracy metrics improve as $\kappa$ increases.} This is expected since the weights of the behaviour $\GRQN$ are updated more frequently resulting in a more accurate prediction of the $Q$-values associated with each~{$\pomdp$} state $\bS_{i}$.~{For instance, the average regret of Algorithm~1 for $\kappa = 3$ is $765.33\times 10^{3}$ $\mws$, which is $3.04\%$ lower than the average regret when $\kappa = 1$. Similarly, the average precision for $\kappa = 3$ is $0.169$, which is $26\%$ higher than the average precision when $\kappa = 1$. To further assess how the accuracy metrics scale with $s\in[S]$, figures~\ref{fig:Accum Risk} and~\ref{fig:Num Risky} illustrate the accuracy versus $s\in [S]$. The observations are consistent with Table~\ref{table:39 bus}, where it is observed that a higher $\kappa$ results in a lower average regret and greater average precision. It is noteworthy that the advantage of the setting $\kappa=3$ is viable at the expense of incurring a higher computational cost. This is due to more frequent weight updates rendering an inevitable accuracy-complexity trade-off.}

\subsubsection{Comparison with Baselines} 
\label{sec:Comparison with Baselines}

To further illustrate the merits of the proposed graphical framework, we compare it with two state-of-the-art baseline approaches proposed in~\cite{Zhang:2020}. We label the first approach in~\cite{Zhang:2020} based on the ordinary $Q$-learning algorithm without prior knowledge as $\PFW$ + $\rl$ and label their best performing approach based on transition and extension of prior knowledge from other power system snapshots by $\PFW + \rl + \TE$. To ensure a fair comparison, we employ the~\emph{same} exploration schedule for $\epsilon$ discussed in Section~\ref{sec:Fault Chain Search Strategy} with the same parameters and the same discount factor $\gamma$ for all the approaches. Note that, in the $\PFW+\rl+\TE$ approach, we first run their proposed $Q$-learning based approach offline, for a loading condition of $0.6\times{\sf base\_load}$ (bringing in the prior knowledge). This is run for $S = 5000$ iterations to ensure the convergence of their $Q$-learning algorithm. Subsequently, we store its extensive $Q$-table to run its $\PFW+\rl+\TE$ approach in real-time for the considered loading condition of $0.55\times{\sf base\_load}$, signifying a transition from the power system snapshot loaded at $0.6\times{\sf base\_load}$ and an extension to the current power system snapshot loaded at $0.55\times{\sf base\_load}$. Note that, when performing comparisons, we set the parameters and hyper-parameters associated with Algorithm~1 the same as that described in Section~\ref{sec:Parameters and Hyper-parameters}.

\begin{table*}[t]
\centering
\scalebox{0.59}{
	\begin{tabular}{| c | c | c | c | c | c |} 			
		\hline
		& \multicolumn{3}{c|}{ \thead{Evaluation Metrics} } 
		& \multicolumn{2}{c|}{ \thead{Accuracy Metrics} }\\ [2.0ex]
		\hline
		\thead{Algorithm}     
		& \thead{Average No. of $\fc$ \\ Sequences $S$ Discovered}
		& \thead{Range for Accumulative $\tll$\\ $\sum_{i=1}^{S} \tll(\mcV_{i})$ (in $\mws$)}   
		&  \thead{Range for the No. of Risky\\ $\fcs$ $\sum_{i=1}^{S}\mathds{1}( \tll(\mcV_{i}) \geq M )$}  
		& \thead{Range for} ${\sf Regret}(S)$ (in $\mws$)
		& \thead{Range for} ${\sf Precision}(S)$ \\ [2.0ex] 
		\hline
		
		Algorithm~1$\;(\kappa = 3)$  & $575$  & $63.484\times 10^{3}\;\pm\;36.7\%$  & $104\;\pm\;33$    & $457.09\times 10^{3}\;\pm\;5.9\%$   & $0.1791\;\pm\;31\%$   \\ [1.0ex] 
		\hline
		
		Algorithm~1$\;(\kappa = 2)$  & $700$  & $79.792\times 10^{3}\;\pm\;39\%$  & $123\;\pm\;38$  & $521.47\times 10^{3}\;\pm\;7.5\%$    & $0.1752\;\pm\;31.5\%$   \\ [1.0ex] 
		\hline
		
		Algorithm~1$\;(\kappa = 1)$  & $937$  & $70.848\times 10^{3}\;\pm\;33.3\%$  & $117\;\pm\;34$  & $673.56\times 10^{3}\;\pm\;4.76\%$   & $0.1249\;\pm\;29.1\%$   \\ [1.0ex] 
		\hline
	    
	    $\PFW$ + $\rl$ + $\TE$ \cite{Zhang:2020} & $1608$ & $68.74\times 10^{3}\;\pm\;3.4\%$  & $112\;\pm\;4$ & $965.41\times 10^{3}\;\pm\;1.61\%$  & $0.0698\;\pm\;2.91\%$   \\[1.0ex] 
		\hline
		
		$\PFW$ + $\rl$ \cite{Zhang:2020} & $1611$  & $65.35\times 10^{3}\;\pm\;3.9\%$    & $105\;\pm\;5$  & $969.78\times 10^{3}\;\pm\;2.27\%$ & $0.0653\;\pm\;5.07\%$   \\ [1.0ex] 
		\hline
	\end{tabular}%
	}
	\caption{{Performance comparison for a computational time of $5$ minutes for the IEEE-$39$ New England test system.}}
	\label{table:time_39 bus}
\end{table*}

\paragraph{Comparison under Unbounded Computational Budget} 

\begin{figure}[t]
 	\centering
 	\includegraphics[width=0.65\linewidth]{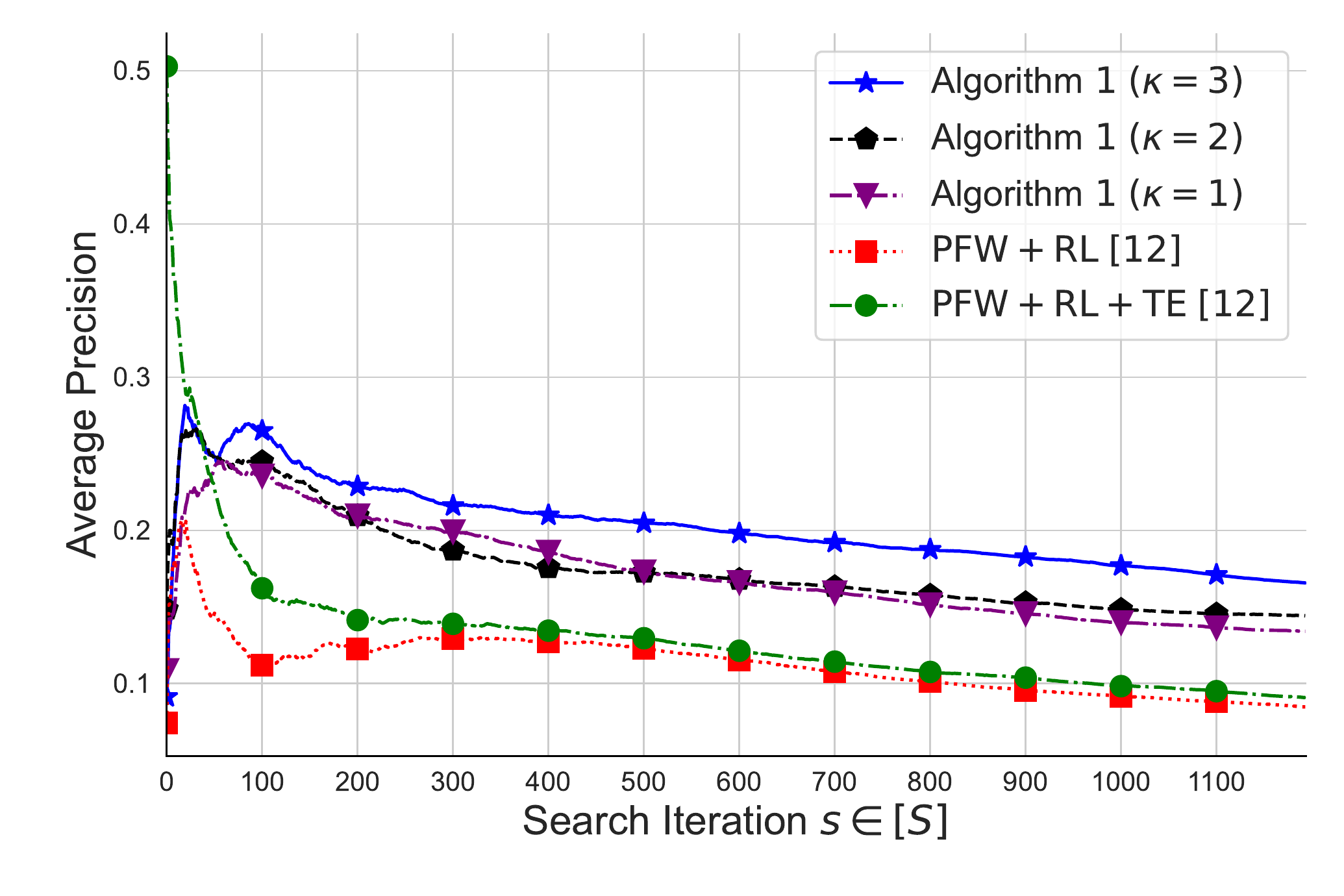}
 	\caption{{${\sf Precision}(s)$ versus $s$ for the IEEE-$39$ bus system.}}
 	\label{fig:Num Risky}
 	\centering
\end{figure}

In parallel to Algorithm~1, we simultaneously run the $Q$-learning update discussed in~\cite{Zhang:2020} for both the $\PFW+\rl$ and $\PFW+\rl+\TE$ approaches to perform comparisons. Table~\ref{table:39 bus} compares the accuracy metrics~{for $S=1200$}, showing that Algorithm~1 consistently outperforms both the other baseline approaches by a wide margin.~{For instance, Algorithm~1 with $\kappa=3$ renders an average regret that is $6.2\%$ smaller than the regret associated with the best preforming baseline $\PFW$ + $\rl$ + $\TE$ approach.} Furthermore, we have two more key observations. ~{First, Algorithm~1 with $\kappa=3$ finds $\fc$ sequences whose accumulated $\tll$ is almost double than that of the two baseline approaches. 

Secondly, even though our approach is designed to optimize the accumulated $\tll$s~\eqref{eq:OBJ1} that is quantified via regret~\eqref{eq:accuracy-1}, it also finds a larger number of risky $\fcs$ (on average). This is reflected in Table~\ref{table:39 bus} and Fig.~\ref{fig:Num Risky}. It is noteworthy  that the baseline approach $\PFW$ + $\rl$ + $\TE$~\emph{outperforms} Algorithm~1 for the first fifty search iterations as shown in Fig.~\ref{fig:Num Risky}. This is expected since our approach does not assume any prior knowledge of the failure dynamics while the $\PFW$ + $\rl$ + $\TE$ approach brings in prior knowledge via the extensive $Q$-table computed offline for the power system snapshot loaded at $0.6\times{\sf base\_load}$. However, as $S$ increases, Algorithm~1 learns the failure dynamics more accurately, resulting in improved accuracy metrics than that of baseline approaches. For example, the average precision of Algorithm~1 with $\kappa=3$ is roughly $86\%$ more than that of $\PFW$ + $\rl$ + $\TE$ and $99\%$ more than $\PFW$ + $\rl$.}

\paragraph{Comparison under Bounded Computational Budget} 

The previous subsection focused on the performance, sans the computational complexity of performing each $\fc$ sequence iteration $s\in[S]$. For real-time implementation, the computational complexity of the $\fc$ search should be within the period of a dispatch cycle. Hence, we evaluate all the above approaches considering a given computational budget. Specifically, we consider the same evaluation and accuracy metrics as discussed in Section~\ref{sec:Accuracy Metrics and Evaluation Criteria} and evaluate them considering a strict run-time of five minutes for the algorithms, averaged over 50 Monte Carlo iterations. Table~\ref{table:time_39 bus} illustrates the relative performance of different algorithms under budget.~{There are three main observations. First, within the $5$ minute computational time budget, for $\kappa = 3$, the number of $\fc$ sequences discovered is considerably smaller than other algorithms. This is reflected in the second column of Table~\ref{table:time_39 bus}. This observation is expected since a large value of $\kappa$ necessitates a larger computation time per $\fc$ search iteration due to the gradient update~\eqref{eq:gradient update}, and hence, affords fewer search iterations. Secondly, when comparing the evaluation metrics, Algorithm~1 with $\kappa = 2$ finds the greatest accumulated $\tll$ and also the largest number of risky $\fcs$, on average. Although both $\PFW$ + $\rl$ and $\PFW$ + $\rl$ + $\TE$ approaches find the most number of $\fc$ sequences, $S=1608$ and $S=1611$, respectively, the quality of the $\fcs$ found are inferior compare to Algorithm~1 with $\kappa = 2$ since their accumulative $\tll$ and the number of risky $\fcs$ are smaller. Third, when comparing the accuracy metrics, the average regret and precision for Algorithm~1 with $\kappa = 3$ outperform other algorithms. Although this approach only finds $S=575$ $\fcs$ on average, it yields the most quality $\fc$ sequences since the frequent update of the behavior $\GRQN$ weights results in a more accurate $Q$-value prediction that optimize~\eqref{eq:OBJ1} better~\emph{per} search iteration $s\in[S]$. Compared to non-graphical algorithm counterparts, although it can find fewer $\fc$ sequences $S$, it has been able to identify the more relevant sequences of interest.} 

It is noteworthy that we have evaluated all the algorithms on a standard computer with no Graphics Processing Units (GPUs). By leveraging GPUs, our approach can accelerate the risky $\fc$ search process even further as GPUs are designed to facilitate the operations involving matrix and vectors for efficient training of the $\GRQNs$, as opposed to other two baselines approaches that, cannot be accelerated for a given computation time.

\subsection{{IEEE-118 Test System}}
\label{sec:IEEE-118 test system}

{This test system consists of $N = 118$ buses and $|\mcU| = 179$ components. We consider a loading condition of $0.6 \times {\sf base\_load}$, where ${\sf base\_load}$ denotes the standard load data for the IEEE-$118$ test case in PYPOWER after generation-load balance to quantify the performance of our approach.
}

\begin{table*}[t]
\centering
\scalebox{0.7}{
	\begin{tabular}{| c | c | c | c | c |} 			
		\hline
		& \multicolumn{2}{c|}{ \thead{Evaluation Metrics} } 
		& \multicolumn{2}{c|}{ \thead{Accuracy Metrics} }\\ [2.0ex]
		\hline
		\thead{Algorithm}               
		& \thead{Range for Accumulative $\tll$\\ $\sum_{i=1}^{S} \tll(\mcV_{i})$ (in $\mws$)}   
		&  \thead{Range for the No. of Risky\\ $\fcs$ $\sum_{i=1}^{S}\mathds{1}( \tll(\mcV_{i}) \geq M )$}  
		& \thead{Range for} ${\sf Regret}(S)$ (in $\mws$)
		& \thead{Range for} ${\sf Precision}(S)$ \\ [2.0ex] 
		\hline
		
		Algorithm~1$\;(\kappa = 3)$    & $45.73\times 10^{3}\;\pm\;16\%$  & $302\;\pm\;49$  & $321.90\times 10^{3}\;\pm\;2.3\%$   & $0.19\;\pm\;17\%$    \\ [1.0ex] 
		\hline
		
		Algorithm~1$\;(\kappa = 2)$    & $40.42\times 10^{3}\;\pm\;17\%$   & $254\;\pm\;39$  & $327.22\times 10^{3}\;\pm\;2.4\%$   & $0.16\;\pm\;15\%$    \\ [1.0ex] 
		\hline
		
		Algorithm~1$\;(\kappa = 1)$    & $44.62\times 10^{3}\;\pm\;12\%$   & $274\;\pm\;28$  & $323.02\times 10^{3}\;\pm\;1.7\%$   & $0.17\;\pm\;10\%$   \\ [1.0ex] 
		\hline
	    
	    $\PFW$ + $\rl$ + $\TE$ \cite{Zhang:2020} & $34.23\times 10^{3}\;\pm\;2.7\%$   & $208\;\pm\;6$  & $333.40\times 10^{3}\;\pm\;0.28\%$   & $0.13\;\pm\;3\%$  \\[1.0ex] 
		\hline
		
		$\PFW$ + $\rl$ \cite{Zhang:2020}  & $35.50\times 10^{3}\;\pm\;2.4\%$    & $174\;\pm\;8$  & $332.14\times 10^{3}\;\pm\;0.25\%$   & $0.11\;\pm\;5\%$  \\ [1.0ex] 
		\hline
	\end{tabular}%
	}
	\caption{{Performance comparison for the IEEE-$118$ test system.}}
	\label{table:118 bus}
\end{table*}

\subsubsection{{Parameters and Hyper-parameters}}
\label{sec:Parameters and Hyper-parameters 118}

{We choose $H = G = 48$ hidden and output number of features when computing both the hidden system state~\eqref{eq:GRNN time varying hidden state} and the output system state~\eqref{eq:GRNN output}. We use $K = 3$ graph-shift operations for the graph-filter~\eqref{eq:graph filter}, use both $\rho$ and $\sigma$ as the hyperbolic tangent non-linearity $\sigma = \rho = \tanh$ and the ${\sf ReLU}$ non-linearity for the fully-connected $\nn$ to approximate $Q$-values. For other parameters, we choose $F=1$ since we employ voltage phase angles as the only input system state parameter $f$, choose $\gamma = 0.99$, $\epsilon_{0} = 0.01$, a batch size $B = 32$, ${\sf Explore} = 250$, a risk assessment horizon $P=3$, $\fc$ sequence iteration $S = 1600$ (\emph{excluding} the initial ${\sf Explore}$ iterations), learning rate $\alpha = 0.0005$, and $\kappa \in [3]$. We set $M$ to $5\%$ of total load (where the total load is $0.6\times{\sf base\_load}$).}

\subsubsection{{Accuracy and Efficiency -- Performance Results}}
\label{sec:Accuracy and Efficiency - Performance Results 118}

\begin{figure}[t]
 	\centering
 	\includegraphics[width=0.65\linewidth]{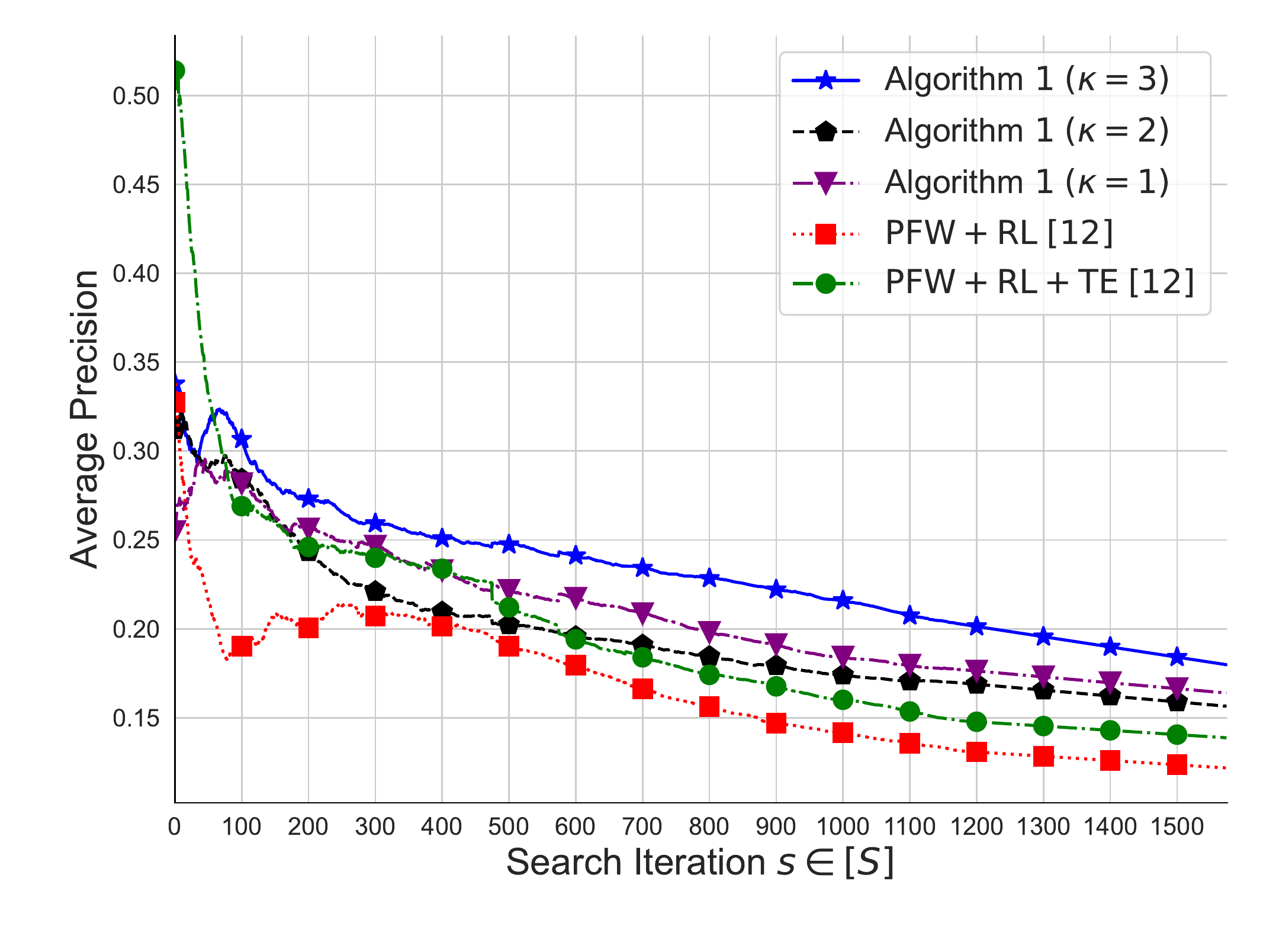}
 	\caption{{${\sf Precision}(s)$ versus $s$ for the IEEE-$118$ bus system.}}
 	\label{fig:Num Risky 118}
 	\centering
\end{figure}

{Algorithm~1 is initiated after the experience buffer is filled for ${\sf Explore}$ search iterations. Table~\ref{table:118 bus} illustrates the results obtained. Similar to $39$-bus system, we observe that Algorithm~1 with a greater $\kappa$ discovers $\fc$ sequences with larger accumulated $\tll$s and discovers more number of risky $\fcs$ leading to superior accuracy metrics for larger $\kappa$. For instance the average regret of Algorithm~1 with $\kappa = 3$ is $321.90 \times 10^{3}$ $\mws$ and it is $0.4 \%$ lower the average regret when $\kappa = 1$. Similarly, the average precision when $\kappa = 3$ is $11.7\%$ higher that of $\kappa = 1$. Figures~\ref{fig:Accum Risk 118} and~\ref{fig:Num Risky 118} illustrate the average accuracy metrics for Algorithm~1 as a function of $s\in[S]$. 
}

\subsubsection{{Comparison with Baselines}} 
\label{sec:Comparison with Baselines 118}

{We perform comparisons with the two baselines approaches in~\cite{Zhang:2020}, i.e., the $\PFW$ + $\rl$ and $\PFW + \rl + \TE$. For the $\PFW+\rl+\TE$ approach, we first run their proposed $ Q$-learning-based approach offline, for a loading condition of $1.0\times{\sf base\_load}$ (bringing in the prior knowledge) for $S = 5000$ iterations and store its extensive $Q$-table to run the $\PFW+\rl+\TE$ approach in real-time for the considered loading condition of $0.6\times{\sf base\_load}$, signifying a transition from $1.0\times{\sf base\_load}$ to an extension to the current system loaded at $0.6\times{\sf base\_load}$. We set the parameters and hyper-parameters as described in Section~\ref{sec:Parameters and Hyper-parameters 118}.}

\paragraph{{Comparison under Unbounded Computational Budget}} 

\begin{figure}[t]
 	\centering
 	\includegraphics[width=0.65\linewidth]{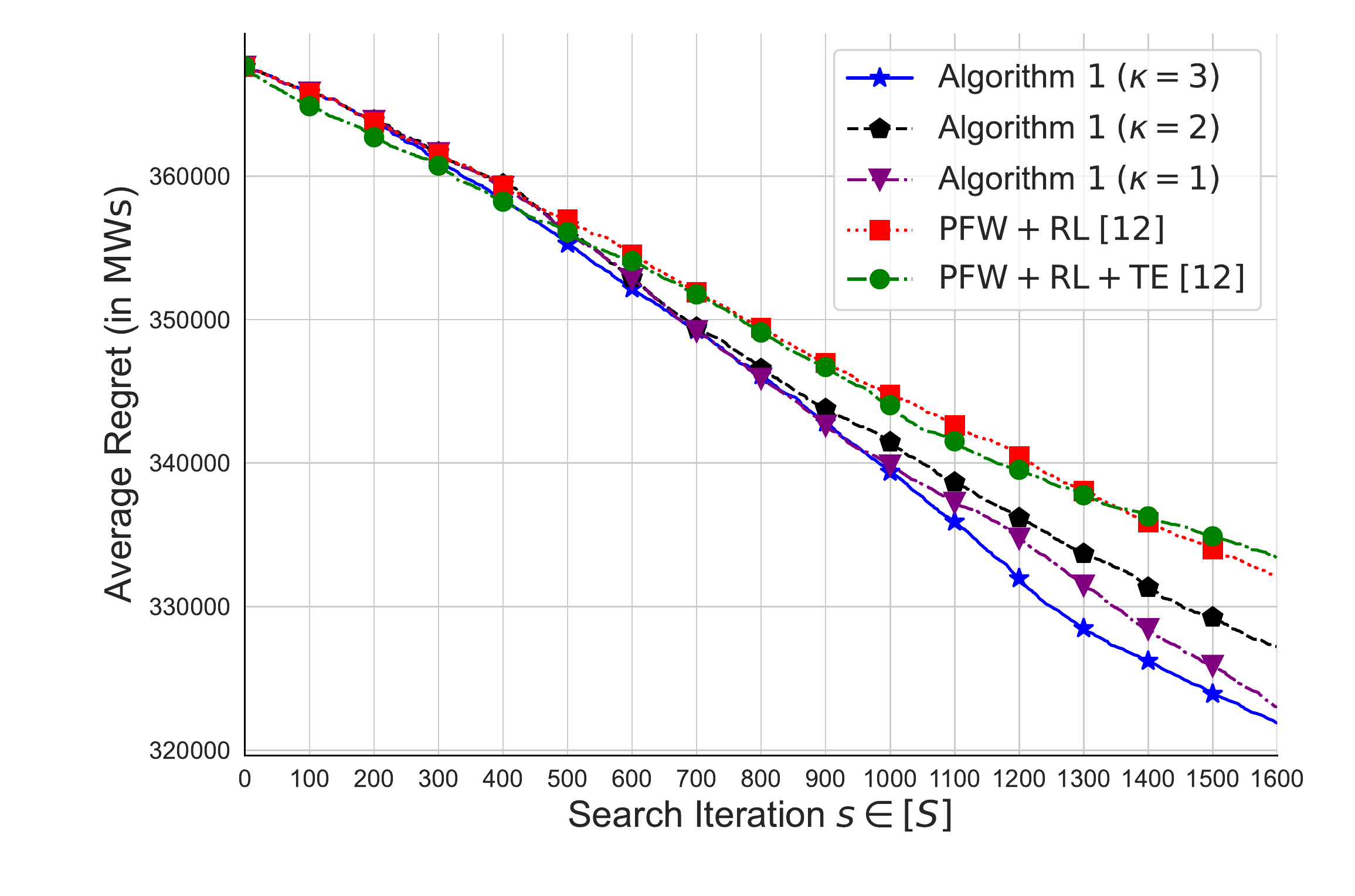}
 	\caption{{${\sf Regret}(s)$ versus $s$ for the IEEE-$118$ bus system.}}
 	\label{fig:Accum Risk 118}
 	\centering
\end{figure}

Table~\ref{table:118 bus} compares the accuracy metrics for $S=1600$, showing that Algorithm~1 consistently outperforms both the other baseline approaches. Figure~\ref{fig:Accum Risk 118} shows how the average regret scales as a function of search iterations $s\in [S]$. Furthermore, even though our approach is designed to optimize the accumulated $\tll$s~\eqref{eq:OBJ1} that is quantified via regret~\eqref{eq:accuracy-1}, it also finds a larger number of risky $\fcs$.

\paragraph{{Comparison under Bounded Computational Budget}} 

We next evaluate all the above approaches considering a run-time computational budget of five minutes, averaged over $25$ $\mc$ iterations, and Table~\ref{table:time_118 bus} illustrates the relative performance. All the observations corroborate those observed for the 39-bus system.

\subsection{{Discussion}}
\label{sec:Discussion}

\subsubsection{Performance Comparisons} The ordinary $Q$-learning algorithm performs only one $Q$-value update for every action taken by the agent due to the intrinsic design of the algorithm illustrated in~\cite{Zhang:2020}. On the other hand, the graph recurrent $Q$-learning algorithm discussed in section~\ref{sec:Graph Recurrent Q learning} can perform multiple gradient updates ($\kappa$ in the inner loop in Algorithm~1) that directly influence the $Q$-values learned by the agent, via the $\GRQN$. This is possible due to the availability of a sequential experience buffer. This, in turn, facilitates learning more efficient strategies in fewer search trials. We also emphasize the approach in~\cite{Zhang:2020} models each permutation of component outages as a unique $\mdp$ state, and as a result, it stores the $Q$-values for a combinatorial number of resulting~{$\mdp$} state-action pairs in an extensive $Q$-table, rendering it not scalable. However, by judiciously leveraging the graphical structure of each $\pomdp$ state and appropriately modeling the dependencies across the various stages of the $\fc$, we have bypassed the storage challenge with fewer modeling assumptions while, at the same time, achieving better performance.

{\subsubsection{Scalability} For large networks, we can further leverage the structure of our $\grnn$ by following an approach based on distributed processing. Specifically, we partition the system into smaller subsystems following the conventional approaches for other monitoring purposes (e.g., estate estimation). The size of subsystems can be decided based on the computational complexity the system operator can afford. The $\fcs$ are subsequently identified within each subsystem. The identified risky $\fcs$ can be subsequently concatenated to form the $\fcs$ for the entire system. This approach, of course, might induce suboptimality in the overall performance of identifying the risky fault chains. Nevertheless, the level of suboptimality induced is expected to be negligible by noting that a fault with a high probability will lead to other faults in its locality.}

\begin{table*}[t]
\centering
\scalebox{0.59}{
	\begin{tabular}{| c | c | c | c | c | c |} 			
		\hline
		& \multicolumn{3}{c|}{ \thead{Evaluation Metrics} } 
		& \multicolumn{2}{c|}{ \thead{Accuracy Metrics} }\\ [2.0ex]
		\hline
		\thead{Algorithm}     
		& \thead{Average No. of $\fc$ \\ Sequences $S$ Discovered}
		& \thead{Range for Accumulative $\tll$\\ $\sum_{i=1}^{S} \tll(\mcV_{i})$ (in $\mws$)}   
		&  \thead{Range for the No. of Risky\\ $\fcs$ $\sum_{i=1}^{S}\mathds{1}( \tll(\mcV_{i}) \geq M )$}  
		& \thead{Range for} ${\sf Regret}(S)$ (in $\mws$)
		& \thead{Range for} ${\sf Precision}(S)$ \\ [2.0ex] 
		\hline
		
		Algorithm~1$\;(\kappa = 3)$  & $186$  & $11.46\times 10^{3}\;\pm\;16.2\%$    & $92\;\pm\;10$   & $175.63\times 10^{3}\;\pm\;3.4\%$ &  $0.201\;\pm\;15\%$\\ [1.0ex] 
		\hline
		
		Algorithm~1$\;(\kappa = 2)$  & $239$  & $18.32\times 10^{3}\;\pm\;18\%$  & $101\;\pm\;12$  & $212.22\times 10^{3}\;\pm\;4.1\%$   & $0.17\;\pm\;14\%$  \\ [1.0ex] 
		\hline
		
		Algorithm~1$\;(\kappa = 1)$  & $301$ & $19.86\times 10^{3}\;\pm\;13\%$  & $115\;\pm\;7$   & $256.75\times 10^{3}\;\pm\;2.6\%$  & $0.18\;\pm\;13\%$ \\ [1.0ex] 
		\hline
	    
	    $\PFW$ + $\rl$ + $\TE$ \cite{Zhang:2020} & $527$ & $18.56\times 10^{3}\;\pm\;3\%$  & $102\;\pm\;3$ & $436.22\times 10^{3}\;\pm\;0.98\%$   & $0.096\;\pm\;5\%$  \\[1.0ex] 
		\hline
		
		$\PFW$ + $\rl$ \cite{Zhang:2020} & $522$  & $17.98\times 10^{3}\;\pm\;3\%$   & $97\;\pm\;4$ & $439.13\times 10^{3}\;\pm\;1.1\%$  & $0.092\;\pm\;6\%$  \\ [1.0ex] 
		\hline
	\end{tabular}%
	}
	\caption{{Performance comparison for a computational time of $5$ minutes for the IEEE-$118$ bus test system.}}
	\label{table:time_118 bus}
\end{table*}

{\subsubsection{Model Adaptation} The computational complexity of learning the failure dynamics can vary across different cascading failure models. The agent's role is to~\emph{learn} the underlying failure dynamics via repeated interactions with the cascading failure simulator. Under different models (e.g., $\dc$ power flow model, $\ac$ power flow model, transient stability model), the complexity of the learning environment changes. As expected, the transient stability-related models are more challenging to learn than $\dc$ power flow-based models under the same number of search iterations $S$. When $S$ is small, the difference in accuracies can be considerable, and the simpler models (e.g., $\dc$ power flow-based) will exhibit better performance. Nevertheless, the performance gap diminishes as the $S$ increases, and the complex models also get a chance to be learned accurately. When prediction accuracies are compared over an arbitrary number of search iterations $S$ (each model can have different search iterations $S$), then we expect all the models to render similar performance since the latent feature representation of the $\grnn$ will be able to better learn the underlying failure dynamics.
}

{\subsubsection{Robustness to Different Cascade Triggers} This paper focuses on triggering mechanisms that fall within the general framework of topological changes, i.e., component failures and their subsequent failures. However, other types of triggering mechanisms such as inappropriate power system control decisions and hidden failures in protection systems, to name a couple, also influence the final load loss. As a result, the type of triggering mechanism directly influences the complexity of the learning problem and the agent decision process. One commonality, however, across the different types of triggering mechanisms is that the underlying topology of the network is bound to change as the cascading failure evolves. Therefore, our framework (which explicitly takes into account the topological changes) can be readily customized to accommodate other triggers. For instance, in cases where power system control decisions trigger the initial failures, the actions space $\mcA_{i}$ can be modeled as continuous, resulting in a more complex agent learning problem. In such cases, an obvious modification would be to augment the input system state $\bX_{i}$ to include more nodal features (e.g., net power injection and voltage magnitudes). This results in a greater amount of information propagated across the hidden layers $\bZ_{i}$. Subsequently, this results in a more complex objective that is a function of the output $\bY_{i}$ of the time-varying $\grnn$ architecture, and the action space $\mcA_{i}$. Optimizing this objective can potentially incur failure paths resulting in maximum load shed. This way, the proposed framework can be robust to the initial fault event type by appropriate re-formulation.}

\section{Conclusion}
\label{sec:Conclusion}

In this paper, we have considered the problem of real-time risky fault chain identification in a limited number of search trials. We have proposed a data-driven graphical framework that can dynamically predict the chains of risky faults a power system faces. First, the search for risky fault chains is modeled as a partially observed Markov decision process. Then a graph recurrent $Q$-learning algorithm is designed to leverage the grid's topology to discover new risky fault chains efficiently. Test results on the IEEE standard systems demonstrate the effectiveness and efficiency of the proposed approach.

\bibliographystyle{IEEEtran}
\bibliography{GRNN}

\end{document}